# Combined fluorescence and photoacoustic imaging of tozuleristide in muscle tissue *in vitro*– toward optically-guided solid tumor surgery: feasibility studies

RUIBO SHANG[1], MATTHEW THOMPSON[2], MATTHEW D. CARSON[3], ERIC J. SEIBEL[3], MATTHEW O'DONNELL[1,*], IVAN PELIVANOV[1,*]

[1]Department of Bioengineering, University of Washington, Seattle, WA 98195, USA
[2]Center for Hip and Knee Replacement, Good Shepherd Orthopedic Surgery, Hermiston, OR 97838, USA
[3] Department of Mechanical Engineering, University of Washington, Seattle, WA 98195, USA

Author Emails: rshang04@uw.edu, mjthompson@gshealth.org, mdc34@uw.edu, eseibel@uw.edu, odonnel@uw.edu, ivanp3@uw.edu

*Correspondence: ivanp3@uw.edu, odonnel@uw.edu, Tel.: +1 (206) 221-8330

# Abstract

Near-infrared fluorescence (NIRF) can deliver high-contrast, video-rate, non-contact imaging of tumor-targeted contrast agents with the potential to guide surgeries excising solid tumors. However, it has been met with skepticism for wide-margin excision due to sensitivity and resolution limitations at depths larger than ~5 mm in tissue. To address this limitation, fast-sweep photoacoustic-ultrasound (PAUS) imaging is proposed to complement NIRF. In an exploratory *in vitro* feasibility study using dark-red bovine muscle tissue, we observed that PAUS scanning can identify tozuleristide, a clinical stage investigational imaging agent, at a concentration of 20 µM from the background at depths of up to ~34 mm, highly extending the capabilities of NIRF alone. The capability of spectroscopic PAUS imaging was tested by direct injection of 20 µM tozuleristide into bovine muscle tissue at a depth of ~ 8 mm. It is shown that laser-fluence compensation and strong clutter suppression enabled by the unique capabilities of the fast-sweep approach greatly improve spectroscopic accuracy and the PA detection limit, and strongly reduce image artifacts. Thus, the combined NIRF-PAUS approach can be promising for comprehensive pre- (with PA) and intra- (with NIRF) operative solid tumor detection and wide-margin excision in optically guided solid tumor surgery.

# Introduction

Solid tumors represent abnormal tissue masses (e.g., sarcomas and carcinomas). Although radiation therapy has improved local control when narrow margins are achieved [1-4], wide excision with negative surgical margins remains the primary treatment [5]. Unexpected positive margins result in worse prognosis [6-9] and carry the greatest chance of local recurrence [1,5]. Inadvertent positive margins in solid tumor (soft tissue sarcoma) were assessed and found to increase risk of local recurrence whether a positive margin was attributable to non-modifiable tumor characteristics or modifiable surgical technique [10]. Subtypes associated with an infiltrative pattern of growth, presumably because the microscopic extent of disease cannot be detected with the naked eye [11], were associated with a greater chance of inadvertent positive margins related to surgical technique [10].

Researchers have pursued advanced imaging techniques for solid tumor detection and margin assessment that perform beyond the naked eye. Among them, near-infrared fluorescence (NIRF) has shown significant promise for oncologic surgeons to localize malignant tissue and assess margin status for a variety of malignancies with high spatial resolution [12-23]. Specifically, fluorescence-guided surgery has shown promise in pre-clinical animal studies to identify solid tumors [14,15,17,20,24]. However, NIRF imaging usually requires line-of-sight since its spatial resolution can be degraded significantly at deep tissue regions due to light scattering [25], restricting potential applications. For example, small amounts of blood or tissue have historically limited NIRF detection, thereby reducing its utility in an ideal solid tumor resection (with ample marginal tissue) - potentially limited only to circumstances where a close margin is unavoidable. Although NIRF can detect dyes 1-2 cm below some tissue surfaces [26], recent quantitative measurements using dual NIRF wavelengths show tissue depth penetrance of ~4 mm [27]. To achieve ~1 cm margins, novel approaches to NIRF detection using fluorescence molecular probes and surgical guidance with minimal levels of target-to-background ratios are being explored to improve subsurface detection in solid tumors [28].

Broad application of fluorescence-guided surgery requires larger (≥ 5 mm) penetration depths. For guidance at greater depths, photoacoustic (PA) measurements can potentially complement fluorescence-guided surgery [29-31]. In PA imaging [32], a pulsed laser illuminates biological tissue. Due to endogenous (e.g., tissue chromophores) or exogenous (e.g., contrast agents) absorbers, biological tissue selectively absorbs laser energy. The nonradiative part of absorption produces local heat expansion that can excite acoustic waves (i.e., PA signals). PA signals can be detected at the tissue surface with conventional [33] or optimized [34,35] ultrasound (US) imaging array transducers. In PA image reconstruction, the laser-induced distribution of heat release is reconstructed from detected PA signals (i.e., acoustic inverse problem). Further signal processing may reconstruct an image of optical absorption at a specific laser wavelength over the region of interest by compensating for the known (or estimated) optical fluence (optical inverse problem) [32]. The reconstruction can be repeated for multiple wavelengths resulting in spectroscopic PA imaging that can distinguish tissues labeled with molecular probes from surrounding tissue constituents based on the difference between the absorption spectra of endogenous chromophores (mainly driven by the blood absorption spectrum) and that of the delivered molecular probe (i.e., contrast agent) [36]. In addition, PA imaging is often co-registered with US imaging because both modalities can share the same detectors.

Hybrid photoacoustic-ultrasound (PAUS) systems using handheld US arrays have several limitations compared to single-modality PA imaging [37-41], but also some advantages. First, PA complements clinical or laboratory US scanners, where US is the primary modality providing high quality anatomic images as well as derivative scans such as color flow images and elastograms. In Jeng et al. (2021) [32], a unique fast-sweep PAUS system was demonstrated. It used a portable, high end, high repetition rate (up to 1 kHz), diode-pumped, wavelength-tunable laser for PA signal excitation, enabling full integration of the PA modality into an US scanning protocol. In addition, the fast-sweep approach reduced the overall cost of the PAUS scanner and provided simultaneous real-time compensation of tissue motion and laser fluence correction. Thus, reliable pixel-based information on tissue molecular content within the imaging volume could be obtained. The same PAUS imaging system was used in this study.

The PAUS imaging system complements the NIRF multimodal scanning fiber endoscope [42,43], a novel medical imaging device employed in various vascular beds as a form of angioscopy, as well as in tracts and duct systems for endoluminal imaging. Owing to its miniaturized form, high resolution, and flexibility, it has successfully imaged across a wide range of diagnostic applications.

To combine NIRF and PA imaging, choosing an appropriate molecular probe to detect a heterogeneous collection of solid tumor subtypes is challenging, often requiring multiple agents and possibly dual agent detection. Tozuleristide (Blaze Bioscience, Seattle, WA) is a synthetically produced, cysteine-dense peptide conjugated to indocyanine green (ICG) [44-46]. The peptide from which tozuleristide is derived has affinities for matrix metalloproteinase-2 and annexin A2, which are expressed in a variety of solid tumors [46]. Researchers have applied it to varying fluorescence studies and shown its promise in fluorescence guided identification of solid tumors [14,47,48] (Figure 1). In a preclinical proof of concept, Fidel et al. showed that tozuleristide provides fluorescence contrast when administered before surgery to dogs with naturally occurring spontaneous tumors [14]. Yamada et al. conducted the first-in-human study of tozuleristide to demonstrate its safety and tolerability for fluorescence imaging of skin tumors [49]. Dintzis et al. conducted patient studies to show that tozuleristide can detect tumor tissue in fresh pathology specimens of 23 patients with breast cancer [50]. Another advantage of tozuleristide is that it has both radiative (NIRF) and nonradiative (PA) energy transitions following light absorption. Therefore, tozuleristide is a good candidate for combining NIRF and PAUS.

Here, our main goal is to demonstrate that the fast-sweep PAUS approach [32] is well suited for surgical guidance of molecular-targeted solid tumors with tozuleristide, extending the imaging depth of NIRF fiber-optic instruments. In particular, we perform two feasibility studies to explore combined PAUS and NIRF in an *in vitro* bovine muscle model to image mock tozuleristide-targeted tumors up to a few centimeters in depth. PAUS serves as a broad volume probe that can be used both pre- and intra-operatively, with highly sensitive NIRF guiding the wide-margin full tumor excision.

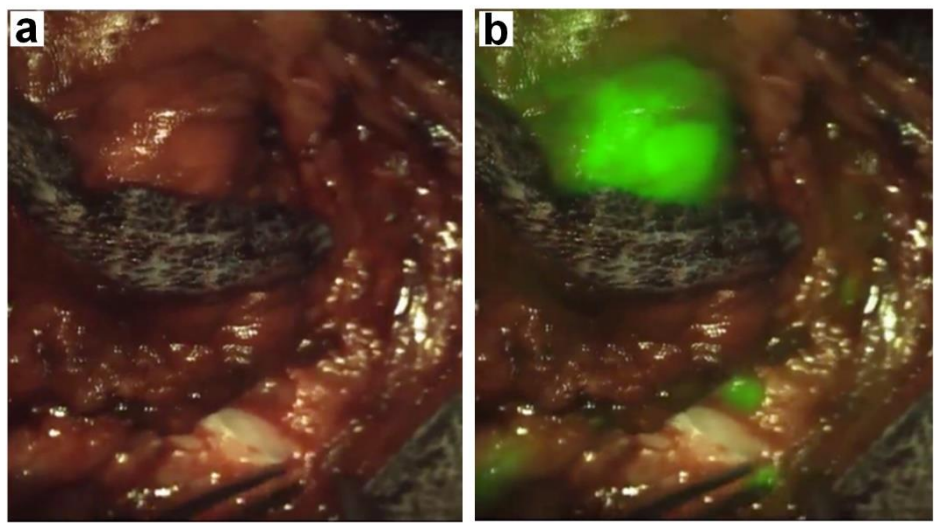

**Figure 1. An example of *in vivo* solid tumor surgery using tozuleristide**. Intraoperative image without (a) and with (b) fluorescence detection during primary resection of solid tumor in a canine administered tozuleristide. Small satellite lesions are detected at the bottom of image (b). Photograph courtesy of Blaze Bioscience, Seattle, WA.

## Methods

No cell lines and laboratory animals were used in this study.

Fresh bovine muscle tissue from an abattoir was used within 48 hours of the animal's death for *in vitro* measurements.

### *Contrast agent - tozuleristide*

Tozuleristide is a clinical-stage, intraoperative fluorescent imaging agent that selectively detects cancerous tissue and can be used in real-time to guide solid tumor resection. It contains a modified form of the chlorotoxin peptide covalently bound to ICG. The natural chlorotoxin peptide was shown to bind tumors via a molecular interaction with protein components of cholesterol-rich lipid rafts, including Annexin A2 and matrix metalloproteinase 2 (MMP2) [51]. In preclinical imaging studies, tozuleristide bound selectively to human-derived and animal tumors, producing observable fluorescence signal and contrast [14,52,53]. It was well-tolerated in IND-enabling toxicology and safety pharmacology studies in rats and non-human primates [54]. The first-in-human clinical study was conducted in patients with skin cancer, noting no dose-limiting toxicities [49]. Other tozuleristide clinical studies include pediatric and adult brain tumors (ClinicalTrials.gov ID: NCT02462629, NCT02234297), breast cancer [50], and other solid tumors (ClinicalTrials.gov ID: NCT02496065). The results from a Phase I adult glioma study were published in 2019 [48].

Tozuleristide has an absorption spectrum in the near infrared as shown in Fig. 2. For NIRF imaging, Tozuleristide is usually excited at 785 nm, with subsequent detection of fluorescent emission between 812 nm and 1000 nm. The light absorption spectrum of tozuleristide is concentration- and matrix-dependent. Light absorption in tozuleristide is then split between radiative (fluorescent emission) and nonradiative (heating) energy transfers. The first is used in NIRF, and the second is used for PA imaging. Thus, the light absorption spectrum, or total absorption, of tozuleristide measured with a UV-VIS spectrophotometer has both radiative and nonradiative components, with an unknown wavelength-dependent split.

In PA imaging, similar to other optical imaging modalities, the concentrations of absorbers are determined by their known absorption spectra using linear unmixing [55] or advanced signal processing [55,56], assuming that a major part of the total absorbance is transferred to heat resulting in PA signals via transient thermal expansion. The accuracy of the unmixing procedure in PA imaging depends on a close match between measured signals and the actual absorption spectrum. It is very important to note that the total absorption spectrum of tozuleristide cannot be used for linear unmixing in PA imaging in the range of optical wavelengths that includes the range of fluorescent transfer since it has a large radiative component. This fact is usually ignored, but as demonstrated in Fig.2, the total absorption of tozuleristide is very different from its nonradiative component, which can lead to serious inaccuracies in spectral unmixing.

To experimentally determine its nonradiative light absorption spectrum, tozuleristide (20 μM concentration in a buffer solution) was sealed in a small transparent tube and immersed in a DI water-diluted milk solution (8× dilution of whole milk) at an 8 mm depth (elevational focal depth of the ultrasound array in the experimental PAUS system) below the US transducer array. Milk was added to DI water for optical scattering. The PAUS system imaged this tube at 35 wavelengths (every 5 nm from 700 nm to 870 nm). Then, the identical tube with 0.4 M cupric sulfate ($CuSO_4 \cdot 5H_2O$) was immersed into the same solution and positioned at the same location (controlled by B-mode ultrasound) as the tube for tozuleristide, and the PAUS system imaged this second tube at the same 35 wavelengths. $CuSO_4 \cdot 5H_2O$ is a molecular absorber with a known, concentration-independent absorption spectrum for concentrations below 1 M [57]. With the known absorption spectrum of 0.4 M $CuSO_4 \cdot 5H_2O$ (measured with UV-VIS spectrophotometry, see Supplementary Note 1) and the ratio of the PA image magnitudes between tozuleristide and $CuSO_4 \cdot 5H_2O$ at the same pixel locations at each wavelength, the nonradiative absorption coefficient of tozuleristide was calculated (Fig. 2, see Supplementary Note 1 for details).

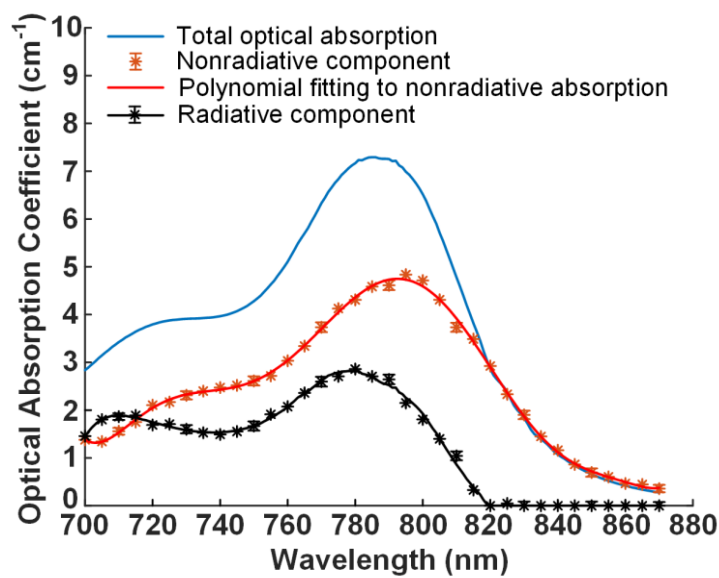

**Figure 2. Optical absorption spectrum of tozuleristide (20 μM concentration in a buffer solution) and its radiative and nonradiative components**. The total optical absorption spectrum of tozuleristide (blue curve) was measured with a UV-VIS spectrophotometer (BioTek, Epoch 2) in the 700 nm - 870 nm range of optical wavelengths with an increment of 1 nm and accuracy of 1%. To determine the nonradiative component (red asterisks) of the tozuleristide spectrum, auxiliary PA measurements in the same wavelength rage were performed with an increment of 5 nm (as described in detail in Supplementary Note 1). A 9$^{th}$ order polynomial fit (red curve) was applied to the calculated nonradiative component to smooth the spectrum. ) The radiative component (black asterisks and red curve) of the tozuleristide optical absorption spectrum was calculated as the difference between its total absorption and the nonradiative absorption.

## *NIRF system*

The multimodal scanning fiber endoscope (mmSFE) jointly developed by the University of Washington and VerAvanti Inc. (illustrated in Fig. 3) was used for NIRF imaging with an excitation laser wavelength of 785 nm, near the peak of tozuleristide's radiative absorption spectrum, that can be mixed with red, green, and blue laser light for multimodal imaging (concurrent NIRF and full color reflectance imaging). Light was delivered through the ultrathin and flexible shaft of the mmSFE by a NIRF single-mode optical fiber. At its distal tip, a piezoelectric ceramic tube was electromechanically vibrated close to the mechanical resonance of the cantilevered fiber, producing a scanned beam as it passes through a microlens system [43,58]. Plastic multimodal optical fibers surround the fiber scanner, collect backscattered reflectance and fluorescence signals, and separate them into four channels using dichroic filters. The NIRF channel (812-1000 nm) used highly sensitive multi-pixel photon counters to produce low-noise NIRF images at a 30 Hz frame rate. NIRF detection was mapped to the green display channel, and all visible lasers were turned off. Although room light was dimmed during all experiments, mmSFE NIRF imaging is not seriously affected by normal room light. In this study, single-frame NIRF imaging was conducted at the surface of fresh red bovine muscle samples for 2D (over the surface) localization of the tozuleristide signal.

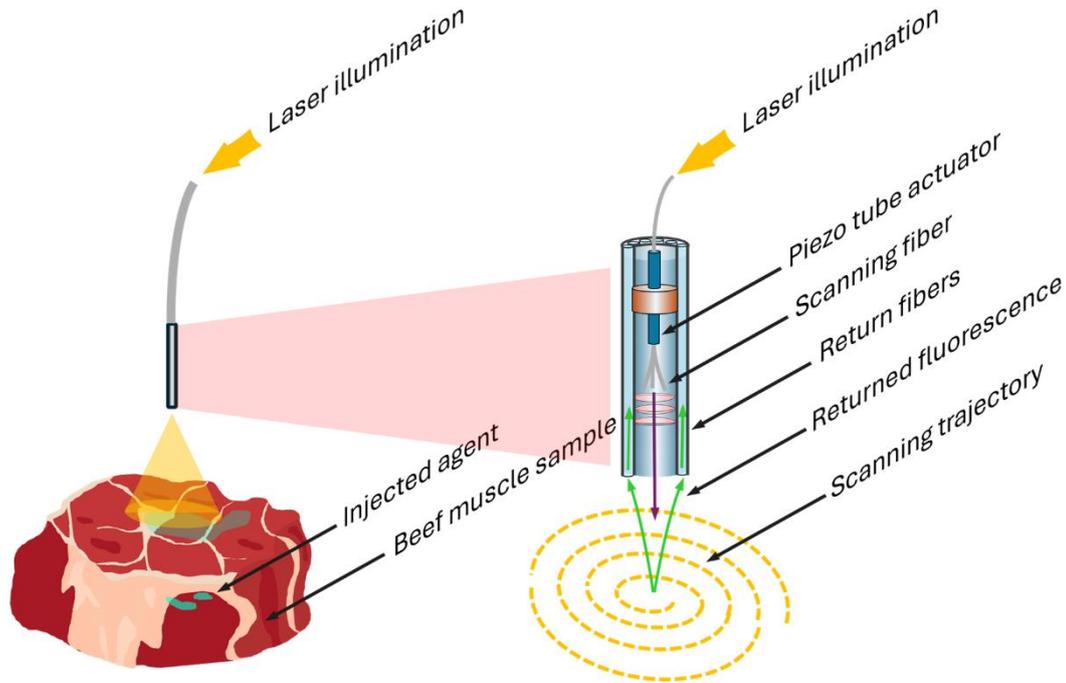

**Figure 3. The multimodal scanning fiber endoscope (mmSFE) for near-infrared fluorescence (NIRF) imaging.** The excitation laser wavelength for this NIRF study was 785 nm, near the peak of tozuleristide's radiative absorption spectrum. The laser beam was delivered to the mmSFE scanning head by a NIRF single-mode optical fiber. At the distal tip of the mmSFE scanning head, the fiber was vibrated using a custom tubular piezoelectric actuator resulting in a cantilever motion of the fiber end that delivered a scanned beam to the sample (*in vitro* red bovine muscle) as it passed through a microlens and created a spiral scanning trajectory consisting of 500 circular scan lines. While scanning the sample, plastic multimodal optical fibers collected backscattered reflectance and fluorescence signals and separated them into four channels using dichroic filters. The entire scanning pattern was repeated at a rate of 30 Hz, providing real-time detection of the internal fluorescent signal emitted from the sample surface.

## *PAUS system*

A lab-built fast-sweep PAUS system (as shown in Fig. 4), described in detail in Jeng et al. (2021) [32], was used for all PA studies. It had an optical fiber delivery system in which each laser pulse was sequentially directed to one of 20 different optical fibers using the position trigger of a motor (TEM Messtechnik GmbH, Germany) spinning with a rate approaching 3000 rpm. At the probe surface, fibers were equally distributed along the long axis of the US array transducer, where 10 fibers were placed along each side. A laser beam from a kHz-rate, wavelength-tunable (700 nm - 900 nm) diode-pumped laser (Laser Export, Russia) was sequentially coupled into the 20 fibers, delivering laser energy to the biological sample for PA imaging. B-mode US signal detection was interleaved with PA acquisition (see Jeng et al. (2021) [32] for details), enabling simultaneous acquisition of PA and US image data at an effective frame rate of about 50 Hz for both US and PA modalities (for a fixed wavelength). The laser wavelength could be tuned from firing to firing for spectroscopic PA imaging in the 700 nm - 900 nm range at a frame rate of $50/N_\lambda$ Hz ($N_\lambda$ is the number of wavelengths). All data acquisition was controlled by a commercial US scanner (Vantage, Verasonics, WA, USA) using trigger signals from the optical delivery system for accurate synchronization between US and PA scan sequences. The fast-sweep optical delivery approach enabled solutions to several key problems in PA imaging, such as real-time wavelength-dependent fluence compensation [32,59] with tissue motion correction [32,59], and strong clutter artifact reduction using compressed averaging [60], as described in Supplementary Note 2.

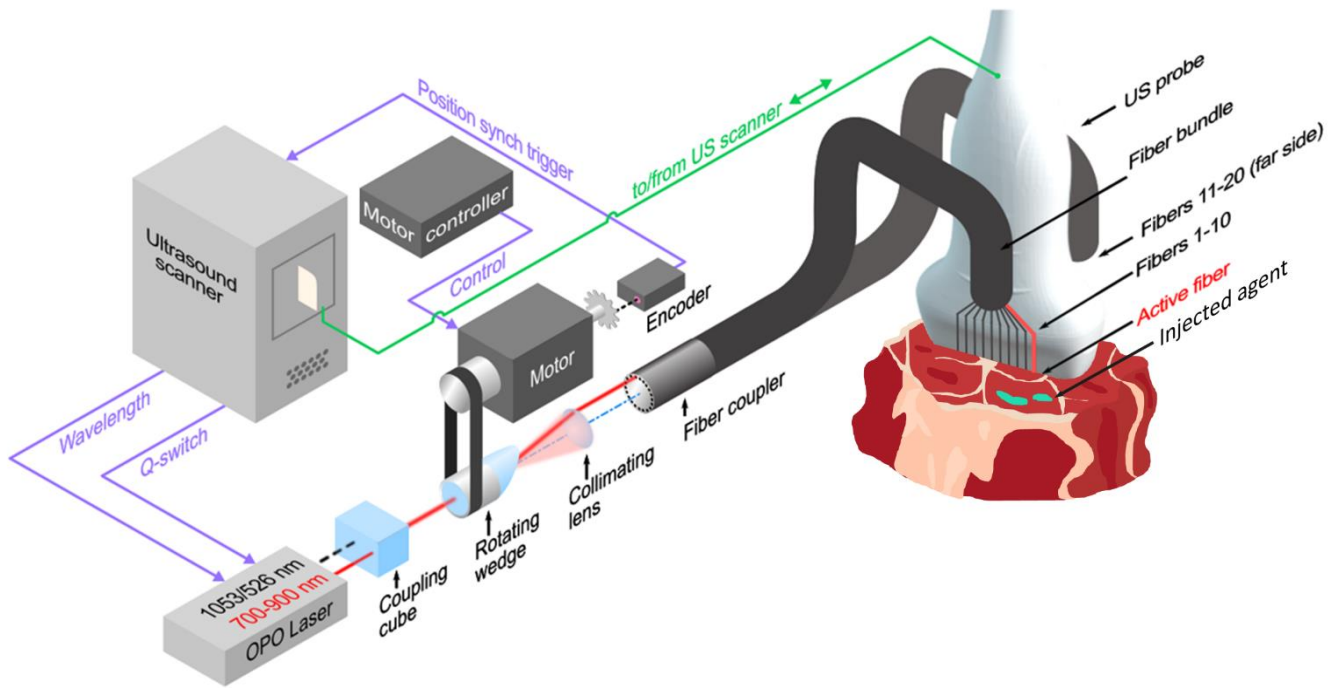

**Figure 4. Fast-sweep photoacoustic-ultrasound (PAUS) imaging system.** The PAUS system consists of a kHz-rate, compact, wavelength-tunable (700 - 900 nm) diode-pumped laser (Laser-Export, Russia), an integrated fiber delivery system (TEM-Messtechnik, Germany), and an US scanner (Vantage, Verasonics, WA, USA) described in detail in Jeng et al. (2021) [32]. Briefly, the spinning motor rotates a wedge with a rate approaching 3000 rpm, offsetting the beam direction from the optical axis and creating a cone beam pattern. A collimating achromatic lens forms a telecentric optical system by directing single-pulse beams to be parallel to the optical axis and focusing them sequentially into each of 20 optical fibers located near the focal plane of the lens. To synchronize laser pulsing with the motor for precise beam coupling into the fibers, the motor encoder sends a trigger signal to the ultrasound scanner (Vantage, Verasonics) based on the angular position of every fiber, which finally initiates the interleaved US B-mode and PA imaging sequence and sends an external trigger to the laser. Thus, twenty (20) triggers per revolution are produced, one for each fiber. These fibers are equally distributed along the long axis of a 15 MHz, 128-element US linear array probe (Vermon, France) with ten (10) on each side. Both PA and US modes are synchronously combined into an interleaved photoacoustic-ultrasound (PAUS) scan sequence operating at about 50 Hz for a full 20-fiber beam rotation cycle at a single optical wavelength. The effective 50-Hz rate PA image consists of 20 sub-images formed by individual fiber illuminations. Spectroscopic PA imaging is achieved by repeating this process at different wavelengths. For each wavelength, the sequence is repeated without any delay for wavelength switching because it is performed between laser pulses (i.e., in less than 1 ms).

### *Localization of tozuleristide using a combined NIRF-PAUS approach*

A possible future strategy for combined PAUS-NIRF image-guided solid tumor localization and margin excision has three important steps. (i) After tozuleristide administration, PAUS imaging can be used pre-operatively to identify and localize a tozuleristide-targeted tumor and its margins up to a few cm in depth. With the information provided by PAUS, a surgeon can manipulate tissue above the solid tumor so that it will be revealed (i.e., line-of-sight). (ii) Then, non-contact NIRF imaging can be used intra-operatively to guide the surgeon for precise margin excision. Note that the sensitivity of PA imaging is not competitive to NIRF when the object is close to or at the tissue surface. Finally, (iii) PAUS imaging can be used *in situ* to confirm that all tozuleristide-targeted tissue has been excised.

In this paper, we investigated the feasibility of step (i) with two studies, as shown in Fig.5. In Study 1 (Figs. 5a-d), both PAUS and NIRF imaging were performed to localize a piece of thin optical lens cleaning tissue (MC-5, Thorlabs, USA) soaked with 20 µM tozuleristide solution and subsequently inserted into a fresh bovine muscle tissue sample at different depths. Ultrasound B-mode images from the PAUS system were used to visualize muscle tissue surrounding the paper insert, and simultaneous PA images localized the insert by identifying the tozuleristide signal, as shown in Figs. 5b and c. For comparison, NIRF imaging was also performed to localize the paper insert as shown in Fig. 5d. Study 1 aimed to demonstrate why PAUS imaging is superior to NIRF imaging to identify and localize a mock tozuleristide-targeted tumor and its margins in step (i). In Study 2, as shown in Fig. 5e, a 20 µM solution of tozuleristide (~300 µl) was directly injected into bovine muscle tissue at a fixed depth, and spectroscopic PAUS imaging was performed to identify only the tozuleristide signal from other PA signals using the image processing steps (detailed in Supplementary Note 2) of fluence compensation, clutter suppression, and tozuleristide-weighted image formation from normalized cross-correlation (NCC) maps. Details and results for the two studies are presented in the next section.

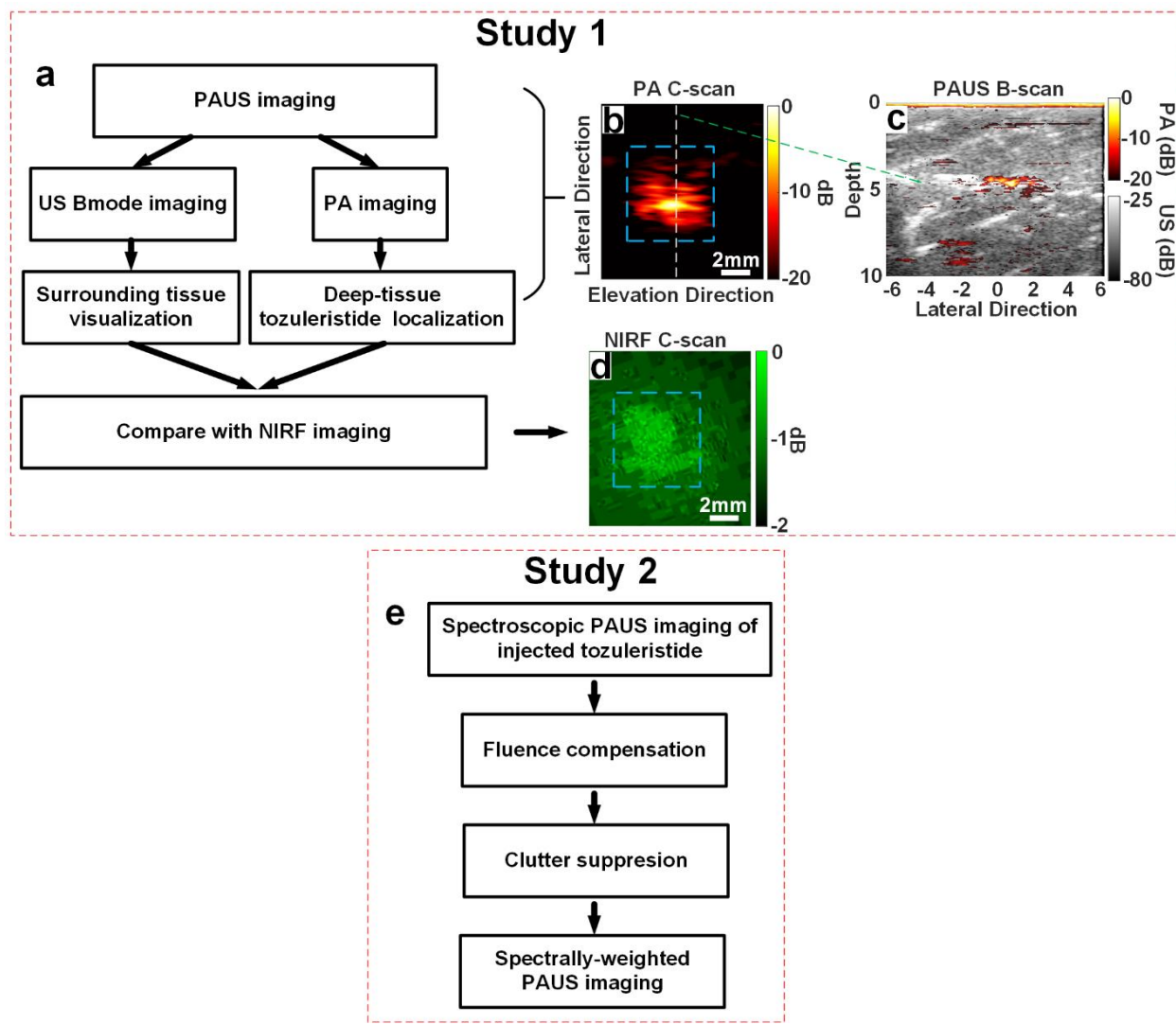

**Figure 5. Description of feasibility studies for the combined PAUS-NIRF approach to localize tozuleristide as a potential strategy for image-guided solid tumor localization and margin excision**. (a-d) Study 1 compares PAUS imaging to NIRF imaging for localizing tozuleristide. A piece of thin optical lens-cleaning-tissue (MC-5, Thorlabs, USA) soaked with 20 μM tozuleristide solution was inserted into a fresh bovine muscle tissue sample at different depths. Then, the sample was imaged with the PAUS system. US B-mode images were used to visualize muscle tissue surrounding the paper insert, and simultaneous PA images localized the paper insert by identifying the tozuleristide signal. The lateral-elevational location of the paper insert is shown in the PA C-scan image in (b), and the depth of the paper insert and surrounding muscle tissue in each column of the PA C-scan image (white dashed line in (b)) is shown in the PAUS B-scan image in (c). For comparison, NIRF imaging (C-scan) was also performed to localize the paper insert as shown in (d). (e) Processing steps of Study 2 using spectroscopic PAUS imaging. A 20 μM solution of tozuleristide was directly injected into bovine muscle tissue at a fixed depth, and spectroscopic PAUS imaging was performed to distinguish the tozuleristide signal from other PA signals.

## Results

### *Study 1: in vitro study of resolution and maximum imaging depth in combined NIRF-PAUS detection of tozuleristide*

The purpose of this study was to evaluate both the maximum imaging depth and resolution of NIRF and PAUS imaging systems, respectively, and to demonstrate why the combined NIRF-PAUS imaging approach can potentially outperform stand-alone NIRF at pre- and intra-operative steps for deep tissue solid tumor localization and margin excision.

Fresh bovine muscle tissue from an abattoir was used within 48 hours of the animal's death for *in vitro* measurements. Each sample was completely immersed in 9g/L saline in a container and then placed in a degassing device to remove internal air bubbles

(until no air bubbles came out of the sample). Then, they were partially divided like an open book (as shown in Fig. 6a) at depths of 4.5 mm, 6.5 mm, 8.8 mm and 14.5 mm from the surface. Pieces of a thin optical lens cleaning tissue (MC-5, Thorlabs, USA) soaked in a 20 µM Tozuleristide solution were subsequently inserted into the tissue samples at the four target depths (as shown in Fig. 6a). Prepared paper samples were 6.7 mm × 6.1 mm, 7.4 mm × 5.5 mm, 6.9 mm × 6.1 mm and 6.0 mm × 4.9 mm at the four depths, respectively. Finally, the top layer of each sample was repositioned to cover the paper inserts. The approach mimicked a significant tumor at different depths from the surface.

In NIRF imaging, the 1.2-mm diameter fiber-optic probe head of the multimodal scanning fiber endoscope was placed ~1.5 cm above the tissue surface and C-scan NIRF images were acquired from the tissue surface as shown in Fig. 6b.

For PAUS imaging, an ultrasound gel was applied to the tissue surface, and the PAUS probe was placed against the surface (as shown in Fig. 6c). PAUS B-scan (both PA and B-mode US) images were obtained (at a fixed wavelength of 795 nm) at one elevational position and a 3D volume of PA data was obtained by recording 19 B-scans at different elevational locations with a step of 0.635 mm between consecutive B-scans. The laser pulse energy was about 0.14 mJ per pulse at the tissue surface, which resulted in about 10.23 mJ/cm$^2$ laser fluence. For each tissue sample, the PA signal at each pixel within a PA B-Scan was averaged over a 1 mm range centered at the depth of the tozuleristide insert to form a PA C-scan image. Finally, acquired NIRF and PAUS images for all four tissue samples were used to evaluate the maximum imaging depth and resolution of NIRF and PAUS imaging systems.

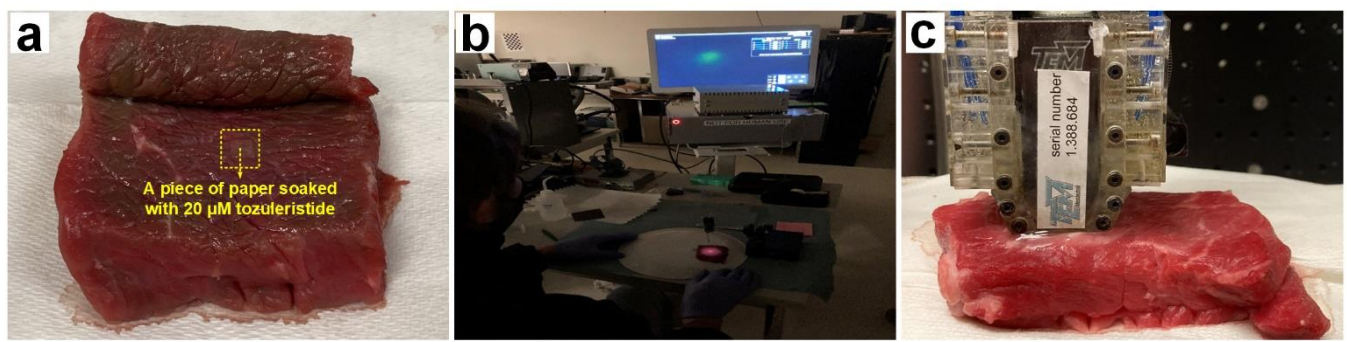

**Figure 6. Study 1 outline.** (a) A piece of lens tissue (MC-5, Thorlabs, USA) was soaked with 20 µM tozuleristide and inserted into a freshly excised (within 48 hours after animal depth) bovine red muscle sample at varying depths from the surface (4.5mm, 6.5mm, 8.8mm and 14.5mm) to mimic a significant tumor at different depths from the surface. Prepared paper samples were 6.7 mm × 6.1 mm, 7.4 mm × 5.5 mm, 6.9 mm × 6.1 mm and 6.0 mm × 4.9 mm at the 4 depths, respectively. The top layer of each sample was then repositioned to cover the paper inserts. *In vitro* NIRF imaging of the muscle sample was performed first. (b) The 1.2-mm diameter fiber-optic probe head of the multimodal scanning fiber endoscope (mmSFE, jointly developed by the University of Washington and VerAvanti Inc.) was placed about 1.5 cm above the tissue surface while monitoring the computer screen and moving the muscle sample to locate the tozuleristide signal. After signal localization, C-scan NIRF images were acquired from the tissue surface. *In vitro* PAUS imaging was then performed on the same muscle sample. (c) The PAUS probe was placed against the sample surface after applying ultrasound gel for acoustic coupling. The excitation laser was tuned to an optical wavelength of 795 nm to match the maximum of tozuleristide's nonradiative absorption. PAUS B-scan images were acquired using the fast-sweep PAUS system at 19 elevational locations with a step of 0.635 mm between consecutive B-scans to produce a 3D volume of PAUS data for subsequent analysis.

Figure 7 shows NIRF and PAUS images acquired in Study 1. The left column of Fig.7 (panels a, d, g) shows C-scans of tozuleristide targets (rectangular lens tissue pieces soaked with 20 µM of tozuleristide) obtained using NIRF *in vitro* at depths of 4.5mm, 6.5mm and 8.8mm. The NIRF image for the 4$^{th}$ tissue sample at a depth of 14.5 mm was not acquired because the tozuleristide signal could not be distinguished from background noise due to high light scattering of the fluorescent signal. Both PA B- (Fig. 7c, f, i and k) and C-scan images (Fig. 7b, e, h and j) are shown for all 4 tissue samples. The PAUS C-scan image was combined from 19 consecutive PAUS B-scan images. Example PAUS B-scan images are shown in Fig. 7c, f, i and k. PA B-scan images show the paper (tozuleristide) signal and its location while the US B-scan images show the surrounding tissue environment. Approximate projections of lens tissue samples soaked in tozuleristide to the tissue surface in all 4 tissue samples are shown by the blue dashed areas in the middle column of Fig.7.

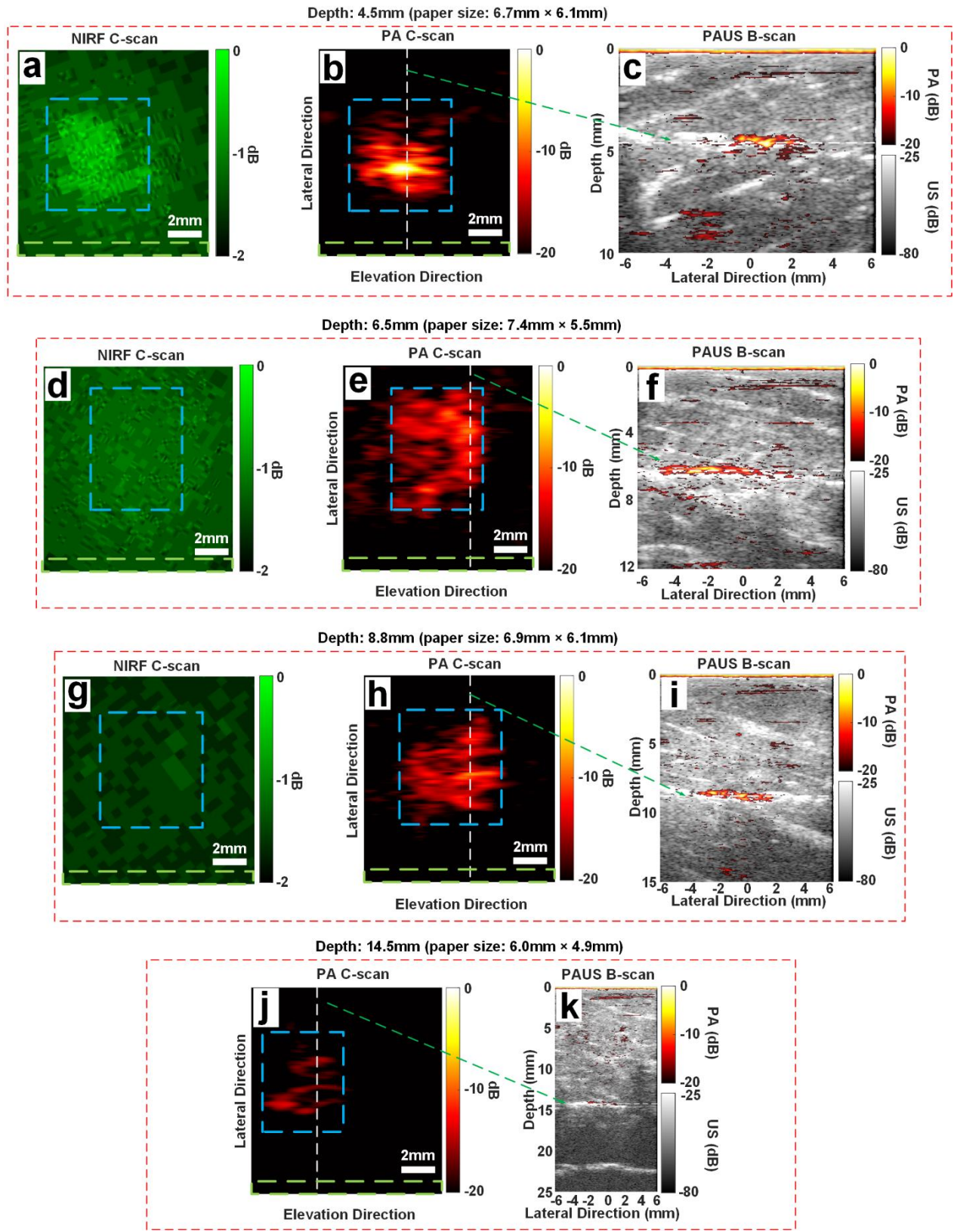

**Figure 7. Comparison of NIRF and PAUS imaging of tozuleristide**. The bovine muscle tissue samples with 20 µM tozuleristide-soaked lens tissue were imaged at different depths (4.5 mm, 6.5 mm, 8.8 mm, and 14.5 mm) using both NIRF and PAUS imaging systems. The multimodal scanning fiber endoscope (mmSFE) was used for NIRF C-scan imaging (a, d, g). The fast-sweep PAUS system was used for PA localization of tozuleristide. PAUS B-scan images were acquired at 19 elevational locations with a step of 0.635 mm between consecutive B-scans to produce a 3D volume of PAUS data for subsequent analysis. PA C-Scan images (b, e, h, j) were obtained as a depth-average over a ~1 mm range of 3D PA data centered at the depth of the tozuleristide insert. Example PA B-Scan images (c, f, i, k) are shown as they were recorded for a fixed elevation position (indicated by a white dashed line in C-scans) for different depths of the tozuleristide insert. In PAUS B-scan images, PA B-scan images show the paper (tozuleristide) signal and its location while the US B-scan images show the surrounding tissue environment. Locations of lens tissue samples soaked in tozuleristide are marked by blue dashed areas.

At an imaging depth of 4.5 mm, both NIRF (Fig. 7a) and PAUS (Fig. 7b,c) results show high sensitivity and spatial resolution, where the boundary of the paper can be clearly identified in both modalities. At a depth of 6.5 mm, the NIRF signal could be slightly differentiated from the background, and the rectangular paper boundary is greatly blurred, showing reduced sensitivity and resolution (Fig. 7d). In contrast, PAUS retains resolution with slightly reduced sensitivity (Fig. 7e, f), and the tozuleristide target and its boundary can be clearly identified. At a depth of 8.8 mm, the NIRF signal could barely be distinguished from the background with greatly reduced sensitivity and resolution (Fig. 7g) whereas PAUS retains resolution with slightly reduced sensitivity (Fig. 7h, i); the contrast agent and its boundary can still be clearly identified. PAUS imaging was also performed at 14.5 mm depth. Although the tozuleristide signal was greatly reduced in the PA image for the target at this depth (Fig. 7k), PA lateral resolution was not sacrificed. Thus, due to light diffusion, the resolution of NIRF was approximately equal to the depth of the target location whereas PA resolution was solely determined by the characteristics of the detecting US array.

To estimate the maximum imaging depth of both NIRF and PAUS modalities, the PA and NIRF signals were analyzed at each depth. First, to help identify primary PA signals in these images, the mean ($\langle Noise\_PA \rangle$) and standard deviation ($\sigma_{Noise\_PA}$) of the background noise were calculated over the dashed green areas in Fig. 7b, e, h and j. To calculate the PA signal at each depth, pixels with PA values larger than $\langle Noise\_PA \rangle + 3\sigma_{Noise\_PA}$ within the corresponding blue dashed area (lens tissue area) were selected. The mean PA value of the selected pixels represents the PA signal at that depth, and the standard deviation of the PA values from the selected pixels represents the standard deviation (error) of the PA signal at that depth (shown as the values and error bars of the data points in Fig. 8a). The same operation was applied to NIRF images to calculate the NIRF signal at each depth (shown as the values and error bars of the data points in Fig. 8b). Note that the primary NIRF signal in Fig. 7g is very close to the NIRF background noise level and, therefore, for Fig. 7g, pixels were selected with NIRF values larger than the mean of the NIRF background noise $\langle Noise\_NIRF \rangle$. Since the signal decreases nearly exponentially with depth, signal points on a dB scale were fit with a simple exponential decay function (a straight line on a dB scale in Fig. 8) to determine the depth at which the signal level was equal to that of background noise. Thus, the maximum imaging depth was estimated to be $34 \pm 9$ mm for PA (Fig. 8a) and $9.6 \pm 0.4$ mm for NIRF modalities (Fig. 8b). For PAUS imaging, the maximum imaging depth can be further increased by optimizing transducer parameters, such as bandwidth and elevational focus. For the current transducer, the elevation focus was 8 mm, which is far from optimal for PA imaging at depths greater than ~ 15 mm. Thus, the estimated maximum imaging depth can be further improved with a properly chosen US probe.

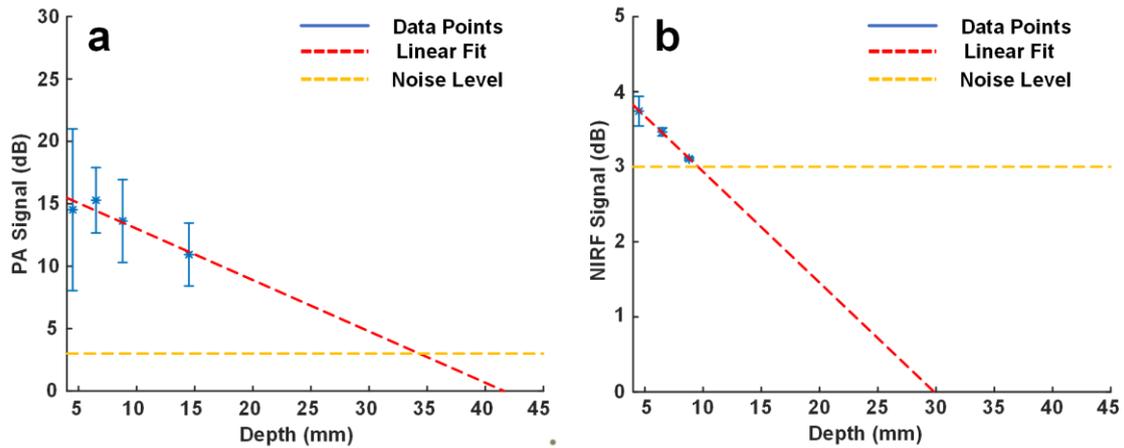

**Figure 8. Depth-dependencies of PA (a) and NIRF (b) signal magnitude.** For PA imaging, the mean and standard deviation values over dashed green areas in Fig. 7b, e, h and j were calculated as the mean ($\langle Noise\_PA \rangle$) and standard deviation ($\sigma_{Noise\_PA}$) of noise in PA C-scan images. To calculate the PA signal at each depth, pixels with PA values larger than $\langle Noise\_PA \rangle + 3\sigma_{Noise\_PA}$ within the corresponding blue dashed area (lens tissue area) were selected and averaged. The standard deviation of the PA signals (error bars in panel (a)) represent point-by-point variations of the PA signals. In the same way, the mean ($\langle Noise\_NIRF \rangle$) and standard deviation ($\sigma_{Noise\_NIRF}$) were computed for NIRF signals over dashed green areas in Fig. 7a, d and g. Dashed Red lines indicate linear fits to experimental points on a dB scale. Yellow lines indicate background noise levels. The linear fit in (a) is $y = a_{PA}x + b_{PA}$ with $a_{PA} = -0.41 \pm 0.11$ dB/mm and $b_{PA} = 17.12 \pm 1.01$ dB. The linear fit in (b) is $y = a_{NIRF}x + b_{NIRF}$ with $a_{NIRF} = -0.15 \pm 0.01$ dB/mm and $b_{NIRF} = 4.41 \pm 0.03$ dB.

***Study 2: in vitro localization of tozuleristide injected into bovine muscle tissue with PAUS***

The purpose of this study was to demonstrate the potential of PAUS to distinguish tozuleristide from intrinsic tissue chromophores using spectroscopic PA imaging, i.e. to demonstrate the accurate measurement of tozuleristide's nonradiative absorption spectrum. A 20 μM solution of tozuleristide was directly injected into bovine muscle tissue at a depth of about 8 mm using a needle (18G – 1 1/2, Becton Dickinson Inc.). Then the needle was removed, and PA imaging was performed at nine different optical wavelengths (730nm, 744nm, 758nm, 772nm, 786nm, 795nm, 814nm, 828nm and 842nm) sequentially. As seen in Fig.9, tozuleristide was mainly distributed along the muscle fiber direction. NIRF imaging was not performed because of its greatly reduced sensitivity at large depths. PA image processing steps included fluence compensation, clutter suppression, and tozuleristide-weighted PA imaging. Signal processing details can be found in Supplementary Note 2. Measurement of the reference nonradiative optical absorption spectrum of the tozuleristide solution was described in **Methods**.

Figure 9 shows PAUS imaging results. Figure 9a shows a wavelength-compounded PA image after fluence compensation for each wavelength. As we reported previously [32], the fast-sweep approach can uniquely separate the contributions of light absorption and laser fluence to the PA pressure magnitude, i.e. solve the fundamental optics problem of PA imaging. In the fast-sweep method, the medium is irradiated sequentially from fibers surrounding the probe, and the resulting image is formed by coherent summation of images formed by single-fiber illumination. Because the single-fiber source moves around the US probe, the distance between the source and target changes for different fibers. This spatial dependence of the individual-fiber PA signal on fiber location can be efficiently used for the direct measurement of laser fluence in the medium, and hence to obtain the fluence-corrected nonradiative optical absorption spectra. The more points used in the PA image for fluence compensation, the higher the accuracy of fluence reconstruction (see details in Supplementary Note 2). In this work, we used all PA image points with magnitudes exceeding 3 times the noise floor, i.e. exceeding $\langle Noise\_PA \rangle + 3\sigma_{Noise\_PA}$. The importance of laser fluence compensation for PA spectroscopic imaging is shown in Fig. 9d, where original and fluence-compensated PA spectra are compared with the true tozuleristide nonradiative absorption spectrum at one image point (indicated by the dashed arrow in Fig. 9). The original (non-compensated) PA spectrum is not accurate and is clearly distorted by the wavelength-dependent laser fluence. Similar dependencies were obtained for other imaging points which contained tozuleristide. The computed optical absorption spectrum can be processed using linear unmixing to identify local tissue chromophores and the presence of contrast agents. Alternatively, if the optical absorption spectrum of the contrast agent is known, the PA spectrum determined for all image points can be correlated to the reference agent spectrum using the normalized cross-correlation coefficient (NCC) to obtain the agent-weighted (tozuleristide-weighted in our case) image (Fig. 9c). Points with high NCC identify the agent with high confidence, as demonstrated in Supplementary Note 2.

In addition to fluence compensation, the fast-sweep approach can be used for efficient clutter suppression. Although the clutter problem is not often discussed, the contribution of the clutter signal can be dramatic in handheld PA imaging. Indeed, deep tissue PA images are formed by strongly attenuated diffuse light. On the other hand, the skin absorbs a large portion of the probe light producing a strong laser-ultrasound signal. This signal propagates into tissue and is scattered by acoustic heterogeneities as in conventional B-mode US, forming artifacts in PA images at double depths. The separation of clutter artifacts from actual PA signals is extremely difficult. However, the fast-sweep approach can efficiently suppress the clutter (see Supplementary Note 2). For clutter suppression, we process single-fiber PA images. Because optical sources move from fiber-to-fiber over the transducer surface, the clutter signal formed by US reflections moves with the optical source whereas the actual PA signal only changes its magnitude (as discussed above). The separation of moving from stationary signals can be performed in different ways, and compressed averaging (see Eqs. S10-S12 in Supplementary Note 2) was used in this work because of its simplicity and ability to be implemented in real time. Figure 9b shows how much cleaner the PA image can be after clutter suppression. We note that more advanced methods can further suppress clutter signals. Figure 9e shows the interleaved PAUS image in which the US B-scan shows tissue anatomy, and the NCC-weighted PA image clearly shows where tozuleristide is located.

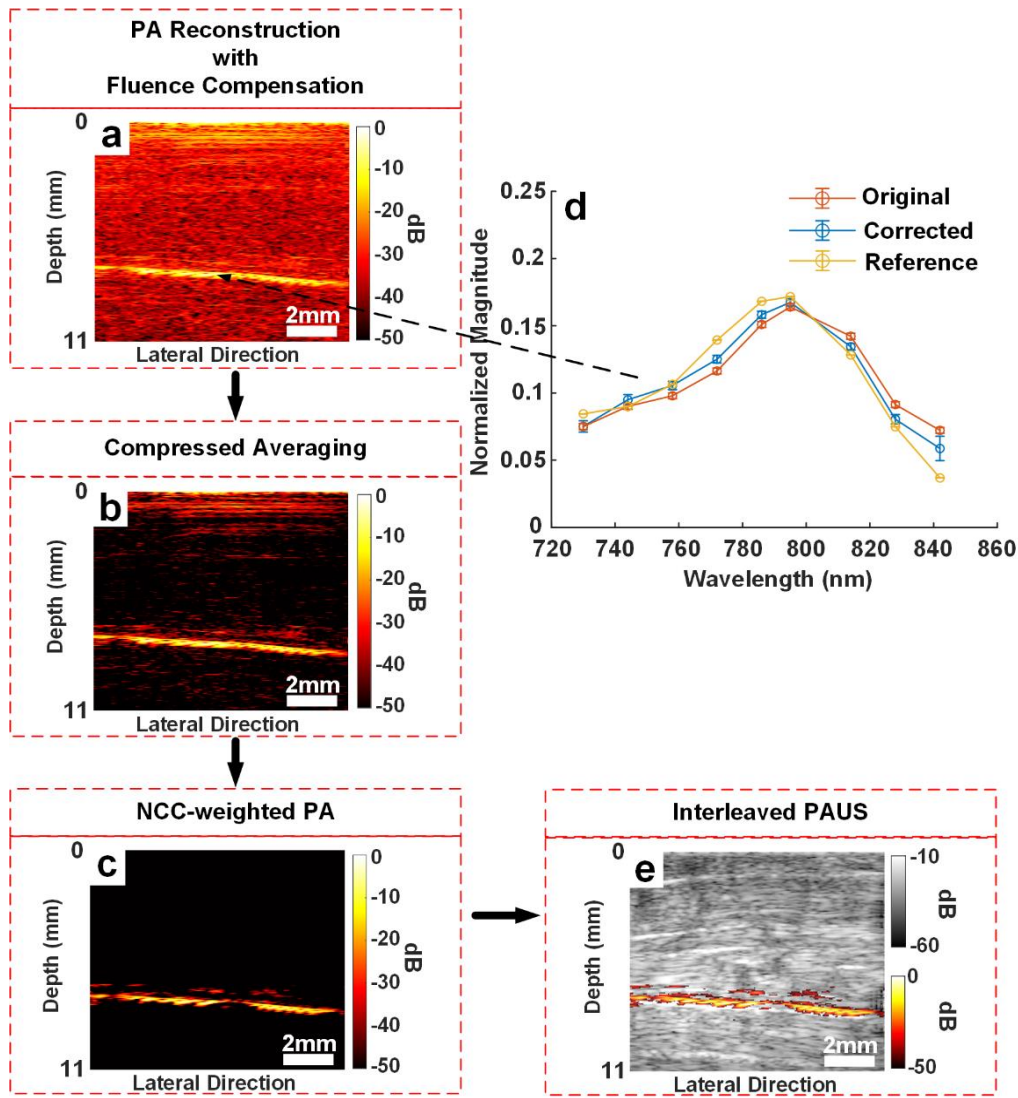

**Figure 9. Spectroscopic PA imaging of tozuleristide injected into bovine muscle tissue.** Signal processing steps included fluence compensation, clutter suppression and NCC-weighting to identify the tozuleristide location. (a) Wavelength-compounded PA image after fluence compensation. Fluence compensation was applied to each pixel at each wavelength using Eq. S9 in Supplementary Note 2, and then fluence-compensated PA images were coherently summed over the investigated wavelength range to form the wavelength-compounded PA image. (b) Wavelength-compounded PA image after fluence compensation and clutter suppression. Fluence compensation was applied to each pixel of the PA image at each wavelength using Eq. S9 in Supplementary Note 2, and then compressed averaging was applied to the fluence-compensated PA images using Eqs. S10-S12 in Supplementary Note 2 before compounding over wavelengths. (c) NCC-weighted PA image (weighted by tozuleristide) after fluence compensation and compressed averaging. The NCC weighting map was calculated using Eqs. S13 and S14 in Supplementary Note 2 and then multiplied with the image in (b). (d) PA-measured nonradiative optical absorption spectrum of tozuleristide at one pixel (indicated by the dashed arrow) within the injection path without (original) and with (corrected) fluence compensation compared to the reference nonradiative absorption spectrum of tozuleristide. Error bars denote the standard deviation of the results. (e) Interleaved PAUS image where the NCC-weighted PA image in (c) is superimposed on the US B-mode image to show both the tozuleristide signal and the surrounding tissue background.

## Discussion

Wide excision of solid tumor with negative margins is the quintessential goal of oncologic surgery. However, factors such as tumor abutment of critical anatomy, infiltrative microscopic tumor growth, or prior non-oncologic excision can make wide excision challenging. Further, as the concept of surgical margins in solid tumor becomes more nuanced, there are clinical scenarios in which narrower margins may result in equivalent oncologic outcomes [1]. In such cases, solid tumors are often successfully treated using current technologies and sound surgical planning, although preoperative imaging may not always accurately predict intraoperative findings [61]. Clearly, there is a clinical need to better guide solid tumor surgeries using non-invasive imaging tools.

NIRF detection of tumors is a powerful tool when a nearly direct line of sight does not compromise outcomes [58]. However, NIRF image sensitivity and resolution for subsurface detection is limited by poor optical penetration, high levels of photon scattering, and background autofluorescence. Thus, the utility of NIRF in wide-margin excision of solid tumors has been limited. However, after endoscopic ablation, the presence of subsquamous intestinal metaplasia (buried Barrett's esophagus) can be found within 3-mm of the surface, so using NIRF probes systemically can extend the range of endoscopic detection of neoplasia using mmSFE.

PA imaging uses non-ionizing laser pulses to stimulate ultrasonic emissions from either endogenous molecules or exogenously delivered contrast agents. It produces images with poorer resolution than NIRF at the tissue surface but yields better image sensitivity and resolution at large penetration depths [62].

Here we used a simple *in vitro* bovine muscle model to quantify subsurface detection of a contrast agent, tozuleristide, using both NIRF and PAUS imaging systems. This agent has been shown to differentially accumulate in solid tumors and is a strong candidate to help guide wide excision treatment. The long-term goal of this work is to further develop both imaging technologies and integrate them into human trials of the surgical treatment of solid tumor.

Both technologies currently require an additional single purpose instrument in the OR and laser safety precautions. As demonstrated here, PAUS enables high-resolution, high-sensitivity detection of tozuleristide in bovine muscle tissue at substantial depths (up to ~34 mm demonstrated here). The maximum depth can be further increased up to ~ 40 mm with optimized laser illumination and transducer selection. We believe that handheld PAUS can be very powerful for preoperative planning of contrast-targeted solid and satellite tumor excision, and to identify targeted tumor intraoperatively after the planned margin excision.

Recent developments in pulsed wavelength-tunable lasers, where diode-pumped, compact, pulse-to-pulse wavelength-tunable OPO lasers are the most impressive, enable full integration of the PA modality into a high-end US clinical scanner to perform spectroscopic PA imaging in near real-time. Spectroscopic PA imaging can be much more accurate in detecting targeted solid tumors compared to single wavelength PA systems. Here we demonstrated the potential of spectroscopic PAUS for this purpose using direct agent injection into muscle. Direct injections may produce bubbles as well, which most likely happened in our case, that can produce PA signals. However, intravenous application of tozuleristide with its subsequent accumulation in the tumor excludes bubbles, but blood in the highly vascularized tumor environment and the products of tumor development may also produce high contrast PA signals. Spectroscopic PA imaging with spectral unmixing or NCC-based agent-weighting makes it possible to sort PA signals and produce high-confidence images of only the agent no matter what the source is of other PA signals.

In contrast, NIRF can be used for immediate feedback intraoperatively since it is non-contact with sensitivity approaching single molecule detection from the tissue surface. Recent advances in the ultra-compact, multimodal scanning fiber endoscope [42,43] make intra-operative NIRF guidance very convenient and robust. As demonstrated here, the PAUS-NIRF approach, which integrates three-modalities - US, PA and NIRF – may provide pre-operative localization with post-operative validation combined with high sensitivity intraoperative guidance. Indeed, US co-registered with PA may help visualize tissue structure and organ morphology for robust surgery planning.

The authors believe that tozuleristide-targeted PAUS-NIRF-guided solid tumor care may optimally and synergistically incorporate both technologies with potential applications in the following contexts:

  - localize small, deep tumors;
  - identify the viable elements of large heterogeneously necrotic tumors for improved diagnostic accuracy of percutaneous tissue biopsy;
  - detect microscopically infiltrative growth, microsatellite lesions and/or regional metastasis;
  - assess margins in real-time, especially when dissecting in previously altered surgical beds or adjacent to critical anatomic structures;
  - plan and execute tumor bed re-excision following surgery;
  - validate outcomes following treatment;
  - target non-surgical ablative therapies.

Although this exploratory collaboration demonstrated strengths and potential applications for both modalities, there are limitations. Our primary goal was to identify synergies between these two optics-based technologies that could be leveraged to develop a single instrument helping to guide solid tumor treatment. Although this feasibility study demonstrated the ability of PAUS and NIRF to localize tozuleristide, future efforts will focus on optimizing sensitivity to clinically relevant concentrations. Due to the massive heterogeneity of solid tumors, it is possible that tumor sub-type may eventually dictate which of a collection of candidate fluorescent agents is most suitable for any given patient, including combinations of agents. Combined with computer assisted surgical navigation [63], next-generation dual-function surgical instruments with subsurface solid tumor detection capabilities could provide real-time guidance in sarcoma surgery with minimal alteration in surgeon workflow.

# Conclusions

To address the limitation of shallow imaging depth for fluorescence-targeted solid tumor visualization, integrated fast-sweep photoacoustic-ultrasound (PAUS) imaging was proposed here to complement NIRF and extend its depth range, enabling improved surgical planning and ensuring visualization of satellite tumors. Using a target containing 20 µM of tozuleristide (a clinical stage investigational imaging agent) located at different depths inside red bovine muscle tissue, we have shown that fast-sweep PAUS can extend NIRF alone up to about 34 mm depth with depth-independent resolution, which can be further extended by optimizing the US transducer design. In addition, we greatly improved the confidence, robustness and signal-to-noise ratio of the PAUS approach by solving a few fundamental problems of handheld PA imaging. First, using multi-fiber sequential tissue illumination, multi-point fluence compensation was demonstrated to faithfully measure the actual absorption spectrum of a target located centimeters inside tissue. This processing can distinguish tozuleristide from endogenous tissue absorbers like blood with high confidence. Second, a realistic real-time approach for laser-ultrasound clutter rejection was proposed and demonstrated. This is critically important for satellite tumor detection because large laser-ultrasound signals are commonly generated at the tissue surface, especially in skin, and superimposed on PA signals potentially creating artifactual satellite tumors. We believe that the combined NIRF-PAUS approach can be promising for comprehensive pre- (with PA) and intra- (with NIRF) operative solid tumor detection and wide-margin excision in optically guided solid tumor surgery.

# Acknowledgements

This research was funded by the National Institutes of Health (NIH) (R01EB030484), with pilot funding from the Anderson Sarcoma Innovation Program at the Fred Hutchinson Cancer Center in Seattle, WA. Funding for the prototype multimodal scanning fiber endoscope (mmSFE) and testing was provided by VerAvanti Inc., through the NCI Direct Phase 2 NIH SBIR Program (R44CA261318). We would like to thank Blaze Bioscience for their in-kind donation of tozuleristide, with special thanks to Julie Novak and Jennifer Johnson. We would like to thank Professor Lilo Pozzo and Maria Politi for their help in measuring the absorption spectrum of tozuleristide. We would like to thank Casey Walsh, MSW, PhD for her assistance with manuscript preparation.

# Author contributions

I.P., M.O.D., M.T. and E.J.S. conceived the idea of combining NIRF and PAUS imaging of tozuleristide in muscle tissue *in vitro*– toward optically-guided solid tumor surgery and supervised this project. R.S., I.P., and M.T. designed the procedures for PAUS imaging of tozuleristide in muscle tissues. R.S., I.P., M.D.C. and E.J.S. conducted the NIRF experiments to image bovine muscle samples with inserted lens cleaning tissues soaked with tozuleristide at varying depths. R.S. conducted the PAUS experiments to image bovine muscle samples with inserted lens cleaning tissues soaked with tozuleristide at varying depths and the experiments to inject tozuleristide into the bovine muscle sample for spectroscopic PAUS imaging, analyzed the results, plotted the figures and wrote the manuscript. All authors revised the manuscript.

# Funding declaration

This research was funded by the National Institutes of Health (NIH) (R01EB030484), with pilot funding from the Anderson Sarcoma Innovation Program at the Fred Hutchinson Cancer Center in Seattle, WA. Funding for the prototype multimodal scanning fiber endoscope (mmSFE) and testing was provided by VerAvanti Inc., through the NCI Direct Phase 2 NIH SBIR Program (R44CA261318).

# Conflict of interest

The concept of dual-purpose surgical dissection instruments with NIRF detection "eyes" is an invention held by the University of Washington, and co-authors M.T. and E.J.S. are participants in a royalty sharing program if commercialized. The other authors declare no conflicts of interest.

# Data and code availability

The dataset and code used in this paper are available upon request.

# References

1. O'Donnell, P. W. et al. The effect of the setting of a positive surgical margin in soft tissue sarcoma. *Cancer* **120**, 2866-2875, doi:10.1002/cncr.28793 (2014).
2. Rosenberg, S. A. et al. The treatment of soft-tissue sarcomas of the extremities: prospective randomized evaluations of (1) limb-sparing surgery plus radiation therapy compared with amputation and (2) the role of adjuvant chemotherapy. *Ann Surg* **196**, 305-315, doi:10.1097/00000658-198209000-00009 (1982).
3. Yang, J. C. et al. Randomized prospective study of the benefit of adjuvant radiation therapy in the treatment of soft tissue sarcomas of the extremity. *Journal of clinical oncology : official journal of the American Society of Clinical Oncology* **16**, 197-203, doi:10.1200/jco.1998.16.1.197 (1998).
4. Pisters, P. W. et al. Long-term results of a prospective randomized trial of adjuvant brachytherapy in soft tissue sarcoma. *Journal of clinical oncology : official journal of the American Society of Clinical Oncology* **14**, 859-868, doi:10.1200/jco.1996.14.3.859 (1996).
5. Müller, D. A., Beltrami, G., Scoccianti, G., Frenos, F. & Capanna, R. Combining limb-sparing surgery with radiation therapy in high-grade soft tissue sarcoma of extremities - Is it effective? *Eur J Surg Oncol* **42**, 1057-1063, doi:10.1016/j.ejso.2016.02.004 (2016).
6. Bell, R. S. et al. The surgical margin in soft-tissue sarcoma. *J Bone Joint Surg Am* **71**, 370-375 (1989).
7. Eilber, F. C. et al. High-grade extremity soft tissue sarcomas: factors predictive of local recurrence and its effect on morbidity and mortality. *Annals of surgery* **237**, 218 (2003).
8. Toulmonde, M. et al. Quality of randomized controlled trials reporting in the treatment of sarcomas. *Journal of clinical oncology : official journal of the American Society of Clinical Oncology* **29**, 1204-1209, doi:10.1200/jco.2010.30.9369 (2011).
9. Fujiwara, T. et al. The adequacy of resection margin for non-infiltrative soft-tissue sarcomas. *Eur J Surg Oncol* **47**, 429-435, doi:10.1016/j.ejso.2020.06.020 (2021).
10. Gundle, K. R. et al. An Analysis of Tumor- and Surgery-Related Factors that Contribute to Inadvertent Positive Margins Following Soft Tissue Sarcoma Resection. *Annals of surgical oncology* **24**, 2137-2144, doi:10.1245/s10434-017-5848-9 (2017).
11. Wada, T. et al. Myxofibrosarcoma with an infiltrative growth pattern: a case report. *Jpn J Clin Oncol* **30**, 458-462, doi:10.1093/jjco/hyd115 (2000).
12. Baik, F. M. et al. Fluorescence Identification of Head and Neck Squamous Cell Carcinoma and High-Risk Oral Dysplasia With BLZ-100, a Chlorotoxin-Indocyanine Green Conjugate. *JAMA Otolaryngol Head Neck Surg* **142**, 330-338, doi:10.1001/jamaoto.2015.3617 (2016).
13. Butte, P. V. et al. Near-infrared imaging of brain tumors using the Tumor Paint BLZ-100 to achieve near-complete resection of brain tumors. *Neurosurg Focus* **36**, E1, doi:10.3171/2013.11.Focus13497 (2014).
14. Fidel, J. et al. Preclinical Validation of the Utility of BLZ-100 in Providing Fluorescence Contrast for Imaging Spontaneous Solid Tumors. *Cancer Res* **75**, 4283-4291, doi:10.1158/0008-5472.Can-15-0471 (2015).
15. Lazarides, A. L. et al. A Fluorescence-Guided Laser Ablation System for Removal of Residual Cancer in a Mouse Model of Soft Tissue Sarcoma. *Theranostics* **6**, 155-166, doi:10.7150/thno.13536 (2016).
16. Guan, G. et al. CXCR4-targeted near-infrared imaging allows detection of orthotopic and metastatic human osteosarcoma in a mouse model. *Sci Rep* **5**, 15244, doi:10.1038/srep15244 (2015).
17. Bartholf DeWitt, S. et al. A Novel Imaging System Distinguishes Neoplastic from Normal Tissue During Resection of Soft Tissue Sarcomas and Mast Cell Tumors in Dogs. *Vet Surg* **45**, 715-722, doi:10.1111/vsu.12487 (2016).
18. Troyan, S. L. et al. The FLARE intraoperative near-infrared fluorescence imaging system: a first-in-human clinical trial in breast cancer sentinel lymph node mapping. *Annals of surgical oncology* **16**, 2943-2952, doi:10.1245/s10434-009-0594-2 (2009).
19. Warram, J. M. et al. Fluorescence imaging to localize head and neck squamous cell carcinoma for enhanced pathological assessment. *J Pathol Clin Res* **2**, 104-112, doi:10.1002/cjp2.40 (2016).
20. Visgauss, J. D., Eward, W. C. & Brigman, B. E. Innovations in Intraoperative Tumor Visualization. *Orthop Clin North Am* **47**, 253-264, doi:10.1016/j.ocl.2015.08.023 (2016).

21  Whitley, M. J. *et al.* A mouse-human phase 1 co-clinical trial of a protease-activated fluorescent probe for imaging cancer. *Sci Transl Med* **8**, 320ra324, doi:10.1126/scitranslmed.aad0293 (2016).

22  de Souza, A. L. *et al.* Fluorescent Affibody Molecule Administered In Vivo at a Microdose Level Labels EGFR Expressing Glioma Tumor Regions. *Mol Imaging Biol* **19**, 41-48, doi:10.1007/s11307-016-0980-7 (2017).

23  Elliott, J. T. *et al.* Simultaneous In Vivo Fluorescent Markers for Perfusion, Protoporphyrin Metabolism, and EGFR Expression for Optically Guided Identification of Orthotopic Glioma. *Clin Cancer Res* **23**, 2203-2212, doi:10.1158/1078-0432.Ccr-16-1400 (2017).

24  Zhu, S., Tian, R., Antaris, A. L., Chen, X. & Dai, H. Near-Infrared-II Molecular Dyes for Cancer Imaging and Surgery. *Adv Mater* **31**, e1900321, doi:10.1002/adma.201900321 (2019).

25  Musnier, B. *et al.* Optimization of spatial resolution and scattering effects for biomedical fluorescence imaging by using sub-regions of the shortwave infrared spectrum. *Journal of Biophotonics* **14**, e202000345 (2021).

26  de Boer, E. *et al.* Optical innovations in surgery. *Br J Surg* **102**, e56-72, doi:10.1002/bjs.9713 (2015).

27  O'Brien, C. M. *et al.* Quantitative tumor depth determination using dual wavelength excitation fluorescence. *Biomedical Optics Express* **13**, 5628-5642 (2022).

28  Samkoe, K. S. *et al.* Application of Fluorescence-Guided Surgery to Subsurface Cancers Requiring Wide Local Excision: Literature Review and Novel Developments Toward Indirect Visualization. *Cancer Control* **25**, 1073274817752332, doi:10.1177/1073274817752332 (2018).

29  Attia, A. B. E. *et al.* A review of clinical photoacoustic imaging: Current and future trends. *Photoacoustics* **16**, 100144, doi:10.1016/j.pacs.2019.100144 (2019).

30  Wissmeyer, G., Pleitez, M. A., Rosenthal, A. & Ntziachristos, V. Looking at sound: optoacoustics with all-optical ultrasound detection. *Light Sci Appl* **7**, 53, doi:10.1038/s41377-018-0036-7 (2018).

31  Stoffels, I. *et al.* Assessment of Nonradioactive Multispectral Optoacoustic Tomographic Imaging With Conventional Lymphoscintigraphic Imaging for Sentinel Lymph Node Biopsy in Melanoma. *JAMA Netw Open* **2**, e199020, doi:10.1001/jamanetworkopen.2019.9020 (2019).

32  Jeng, G. S. *et al.* Real-time interleaved spectroscopic photoacoustic and ultrasound (PAUS) scanning with simultaneous fluence compensation and motion correction. *Nat Commun* **12**, 716, doi:10.1038/s41467-021-20947-5 (2021).

33  Choi, W., Park, E.-Y., Jeon, S. & Kim, C. Clinical photoacoustic imaging platforms. *Biomedical engineering letters* **8**, 139-155 (2018).

34  Manwar, R., Islam, M. T., Ranjbaran, S. M. & Avanaki, K. Transfontanelle photoacoustic imaging: ultrasound transducer selection analysis. *Biomedical Optics Express* **13**, 676-693 (2022).

35  Xia, W. *et al.* An optimized ultrasound detector for photoacoustic breast tomography. *Medical physics* **40**, 032901 (2013).

36  Wang, L. V. *Photoacoustic imaging and spectroscopy*. (CRC press, 2017).

37  Jiang, J., Yuan, C., Zhang, J., Xie, Z. & Xiao, J. Spectroscopic photoacoustic/ultrasound/optical-microscopic multimodal intrarectal endoscopy for detection of centimeter-scale deep lesions. *Frontiers in Bioengineering and Biotechnology* **11**, 1136005 (2023).

38  Oeri, M., Bost, W., Sénégond, N., Tretbar, S. & Fournelle, M. Hybrid photoacoustic/ultrasound tomograph for real-time finger imaging. *Ultrasound in Medicine & Biology* **43**, 2200-2212 (2017).

39  Park, E.-Y., Lee, H., Han, S., Kim, C. & Kim, J. Photoacoustic imaging systems based on clinical ultrasound platform. *Experimental Biology and Medicine* **247**, 551-560 (2022).

40  Kim, M., Pelivanov, I. & O'Donnell, M. Review of deep learning approaches for interleaved photoacoustic and ultrasound (PAUS) imaging. *IEEE Transactions on Ultrasonics, Ferroelectrics, and Frequency Control* (2023).

41  Kim, M., Jeng, G.-S., Pelivanov, I. & O'Donnell, M. Deep-learning image reconstruction for real-time photoacoustic system. *IEEE transactions on medical imaging* **39**, 3379-3390 (2020).

42  Walker, G. *et al.* The Scanning Fiber Endoscope: A Novel Surgical and High-Resolution Imaging Device for Intracranial Neurosurgery. *Operative Neurosurgery* **23**, 326-333 (2022).

43  Lee, C. M., Engelbrecht, C. J., Soper, T. D., Helmchen, F. & Seibel, E. J. Scanning fiber endoscopy with highly flexible, 1 mm catheterscopes for wide-field, full-color imaging. *J Biophotonics* **3**, 385-407, doi:10.1002/jbio.200900087 (2010).

44  Alander, J. T. *et al.* A review of indocyanine green fluorescent imaging in surgery. *Journal of Biomedical Imaging* **2012**, 7-7 (2012).

45 Reinhart, M. B., Huntington, C. R., Blair, L. J., Heniford, B. T. & Augenstein, V. A. Indocyanine green: historical context, current applications, and future considerations. *Surgical innovation* **23**, 166-175 (2016).

46 Hernot, S., van Manen, L., Debie, P., Mieog, J. S. D. & Vahrmeijer, A. L. Latest developments in molecular tracers for fluorescence image-guided cancer surgery. *The lancet oncology* **20**, e354-e367 (2019).

47 Lee, A. *et al.* Phase 1 and expanded imaging study of tozuleristide in patients with pediatric primary central nervous system tumors. *Journal of Neurosurgery: Pediatrics* **1**, 1-11 (2025).

48 Patil, C. G. *et al.* Phase 1 safety, pharmacokinetics, and fluorescence imaging study of tozuleristide (BLZ-100) in adults with newly diagnosed or recurrent gliomas. *Neurosurgery* **85**, E641-E649 (2019).

49 Yamada, M. *et al.* A first-in-human study of BLZ-100 (tozuleristide) demonstrates tolerability and safety in skin cancer patients. *Contemporary Clinical Trials Communications* **23**, 100830 (2021).

50 Dintzis, S. M. *et al.* Real-time visualization of breast carcinoma in pathology specimens from patients receiving fluorescent tumor-marking agent tozuleristide. *Archives of pathology & laboratory medicine* **143**, 1076-1083 (2019).

51 Khanyile, S., Masamba, P., Oyinloye, B. E., Mbatha, L. S. & Kappo, A. P. Current biochemical applications and future prospects of chlorotoxin in cancer diagnostics and therapeutics. *Advanced Pharmaceutical Bulletin* **9**, 510 (2019).

52 Baik, F. M. *et al.* Fluorescence identification of head and neck squamous cell carcinoma and high-risk oral dysplasia with BLZ-100, a chlorotoxin-indocyanine green conjugate. *JAMA otolaryngology–head & neck surgery* **142**, 330-338 (2016).

53 Butte, P. V. *et al.* Near-infrared imaging of brain tumors using the Tumor Paint BLZ-100 to achieve near-complete resection of brain tumors. *Neurosurgical focus* **36**, E1 (2014).

54 Parrish-Novak, J. *et al.* Nonclinical profile of BLZ-100, a tumor-targeting fluorescent imaging agent. *International journal of toxicology* **36**, 104-112 (2017).

55 Li, M., Tang, Y. & Yao, J. Photoacoustic tomography of blood oxygenation: a mini review. *Photoacoustics* **10**, 65-73 (2018).

56 Gröhl, J. *et al.* Learned spectral decoloring enables photoacoustic oximetry. *Scientific reports* **11**, 6565 (2021).

57 Pelivanov, I. *et al.* Molecular fingerprinting of nanoparticles in complex media with non-contact photoacoustics: Beyond the light scattering limit. *Scientific Reports* **8**, 14425 (2018).

58 Chen, J. *et al.* Detection of Barrett's neoplasia with a near-infrared fluorescent heterodimeric peptide. *Endoscopy*, doi:10.1055/a-1801-2406 (2022).

59 Kim, M., Jeng, G.-S., O'Donnell, M. & Pelivanov, I. Correction of wavelength-dependent laser fluence in swept-beam spectroscopic photoacoustic imaging with a hand-held probe. *Photoacoustics* **19**, 100192 (2020).

60 Polichetti, M., Varray, F., Béra, J.-C., Cachard, C. & Nicolas, B. A nonlinear beamformer based on p-th root compression—Application to plane wave ultrasound imaging. *Applied Sciences* **8**, 599 (2018).

61 Harmsen, S., Teraphongphom, N., Tweedle, M. F., Basilion, J. P. & Rosenthal, E. L. Optical Surgical Navigation for Precision in Tumor Resections. *Mol Imaging Biol* **19**, 357-362, doi:10.1007/s11307-017-1054-1 (2017).

62 Peng, D. *et al.* Precise diagnosis in different scenarios using photoacoustic and fluorescence imaging with dual-modality nanoparticles. *Nanoscale* **8**, 14480-14488 (2016).

63 Henderson, E. R. *et al.* Fluorescence guidance improves the accuracy of radiological imaging-guided surgical navigation. *Journal of Surgical Oncology* **127**, 490-500 (2023).

# Supplementary Notes

# Combined fluorescence and photoacoustic imaging of tozuleristide in muscle tissue *in vitro*– toward optically-guided solid tumor surgery: feasibility studies

RUIBO SHANG[1], MATTHEW THOMPSON[2], MATTHEW D. CARSON[3], ERIC J. SEIBEL[3], MATTHEW O'DONNELL[1,*], IVAN PELIVANOV[1,*]

[1]Department of Bioengineering, University of Washington, Seattle, WA 98195, USA
[2]Center for Hip and Knee Replacement, Good Shepherd Orthopedic Surgery, Hermiston, OR 97838, USA
[3]Department of Mechanical Engineering, University of Washington, Seattle, WA 98195, USA

Author Emails: rshang04@uw.edu, mjthompson@gshealth.org, mdc34@uw.edu, eseibel@uw.edu, odonnel@uw.edu, ivanp3@uw.edu

*Correspondence: ivanp3@uw.edu, odonnel@uw.edu, Tel.: +1 (206) 221-8330

# Supplementary Note 1. Measurement of tozuleristide nonradiative absorption spectrum

Since UV-VIS spectrophotometry measurements of the tozuleristide absorption spectrum contain both radiative and nonradiative absorption components, auxiliary experiments were needed to measure its nonradiative absorption. Tozuleristide (20 μM concentration in the buffer solution provided by Blaze Bioscience) was sealed in a small transparent tube as shown in Fig. S1a and immersed in a DI water-diluted milk solution (8× dilution of whole milk) at an 8 mm depth (elevational focal depth of the ultrasound array in the experimental PAUS system) below the US transducer array. Whole milk was added to DI water for optical scattering. The PAUS system imaged this tube at 35 wavelengths (every 5nm from 700nm to 870nm) as shown in Fig. S1c. Then, the identical tube with a 0.4 M cupric sulfate solution ($CuSO_4 \cdot 5H_2O$) as shown in Fig. S1a was immersed into the same DI water-diluted milk solution and positioned at the same location (controlled by B-mode ultrasound) as the tube for tozuleristide, and the PAUS system imaged the second tube at the same 35 wavelengths as shown in Fig. S1c.

The PA magnitudes of tozuleristide and $CuSO_4 \cdot 5H_2O$ calculated from the 3×3 local pixel value average within the yellow box in Fig. S1c (noise level was calculated from a 3×3 local pixel value average within a white box in Fig. S1c, and subtracted from calculated PA magnitudes) are shown in Fig. S1d. Errors of the PA magnitudes were calculated as the standard deviation of the pixel values within the white box regions in Fig. S1c from all 35 wavelengths. The reason to place the $CuSO_4 \cdot 5H_2O$ tube in the same milk solution at the same location as the tozuleristide tube is that the ratio of the PA magnitudes between tozuleristide and $CuSO_4 \cdot 5H_2O$ at the same pixel locations at each wavelength equals the ratio of the nonradiative absorption magnitudes between tozuleristide and $CuSO_4 \cdot 5H_2O$ (because the optical fluence is the same).

Because of the large data volume required for measurements over many wavelengths, real-time beamforming was used in the Verasonics system rather than off-line beamforming of captured data. Note that there are noticeable variations in the normalized PA magnitude across wavelengths caused by different delays at different wavelengths relative to the start of real-time beamforming. The Verasonics beamformer cannot compensate for these wavelength-dependent delays in real-time, which produces small variations in the PA magnitude. However, since these wavelength-dependent delays were the same for tozuleristide and $CuSO_4 \cdot 5H_2O$, they did not affect the ratio of PA magnitudes between them.

$CuSO_4 \cdot 5H_2O$ is a molecular absorber with a known, concentration-independent absorption spectrum for concentrations below 1M [1]. With the known absorption spectrum of 0.4M $CuSO_4 \cdot 5H_2O$ (measured with UV-VIS spectrophotometry as shown in Fig. S1b) and the ratio of the PA magnitudes between tozuleristide and $CuSO_4 \cdot 5H_2O$ at the same pixel locations at each wavelength, the nonradiative absorption coefficients of tozuleristide can be calculated according to Eq. (S1):

$$\alpha_{tozuleristide}(\lambda) = \frac{PA_{tozuleristide}(\lambda)}{PA_{CuSO4 \cdot 5H2O}(\lambda)} \times \alpha_{CuSO4 \cdot 5H2O}(\lambda) \,, \quad (S1)$$

where $\alpha_{tozuleristide}(\lambda)$ is the nonradiative absorption coefficient of tozuleristide at wavelength $\lambda$, $\alpha_{CuSO4 \cdot 5H2O}(\lambda)$ is the nonradiative absorption coefficient of $CuSO_4 \cdot 5H_2O$ at wavelength $\lambda$, $PA_{tozuleristide}(\lambda)$ is the PA magnitude of tozuleristide in the tube at wavelength $\lambda$, and $PA_{CuSO4 \cdot 5H2O}(\lambda)$ is the PA magnitude of tozuleristide in the tube at wavelength $\lambda$. Then, polynomial curve fitting is applied to the nonradiative absorption coefficients of tozuleristide at 35 wavelengths to obtain the coefficients at any wavelength within the 700nm-870nm range.

The wavelength-dependent standard deviation (STD) in the computed nonradiative absorption spectrum $\alpha_{Tozuleristide}(\lambda)$ was estimated as

$$STD(\alpha_{Tozuleristide}(\lambda)) = STD(\frac{PA_{Tozuleristide}(\lambda)}{PA_{CuSO4 \cdot 5H2O}(\lambda)} \times \alpha_{CuSO4 \cdot 5H2O}(\lambda)) =$$

$$= \alpha_{Tozuleristide}(\lambda) \times \sqrt{\left(\frac{STD(PA_{Tozuleristide}(\lambda))}{PA_{Tozuleristide}(\lambda)}\right)^2 + \left(\frac{STD(PA_{CuSO4 \cdot 5H2O}(\lambda))}{PA_{CuSO4 \cdot 5H2O}(\lambda)}\right)^2} \,. \quad (S2)$$

where STD denotes the standard deviation. Other parameters are the same as in Eq. (S1). The standard deviation of the pixel values within the white box regions in Fig. S1(c) from all 35 wavelengths was calculated as the standard deviation of the PA magnitudes of tozuleristide and $CuSO_4 \cdot 5H_2O$.

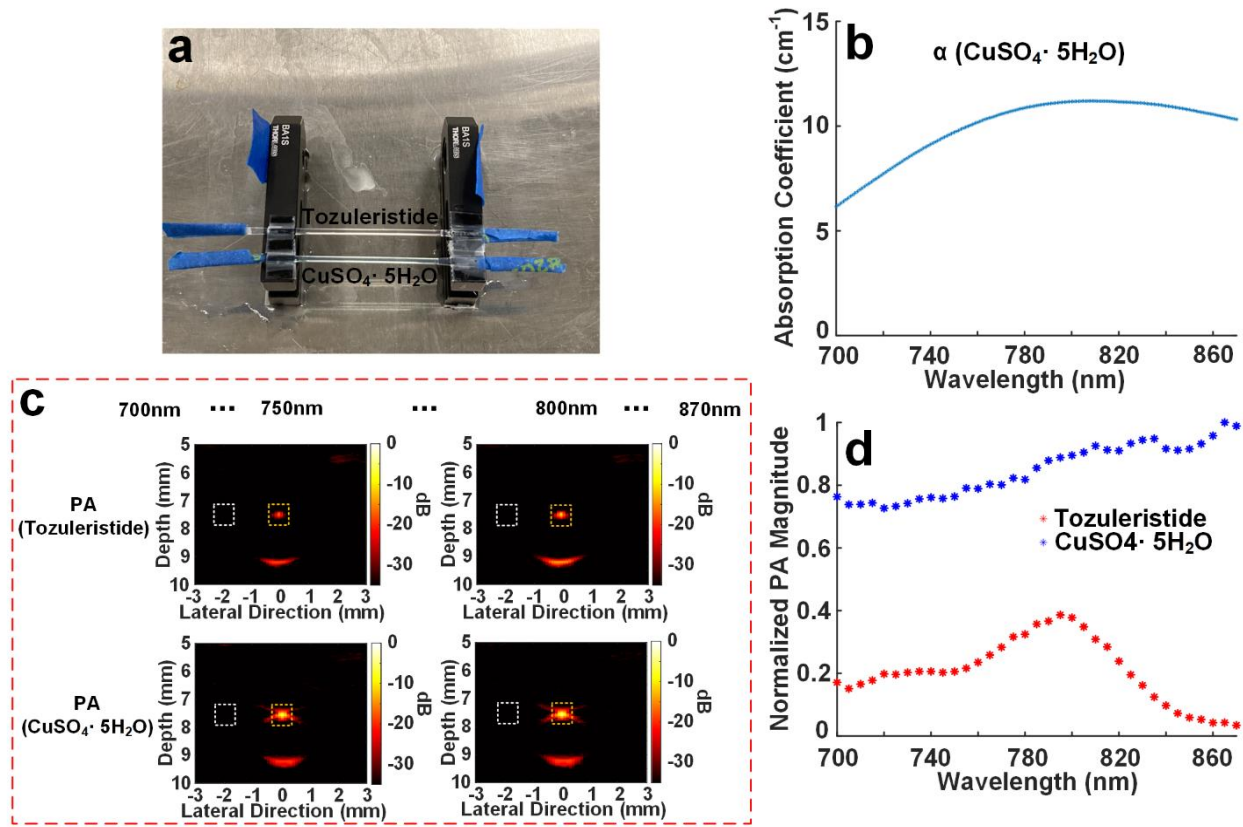

**Supplementary Figure S1.** Tozuleristide nonradiative absorption spectrum measurement. (a) Tubes filled with tozuleristide and $CuSO_4 \cdot 5H_2O$. Tozuleristide (20 µM concentration in a buffer solution) was sealed in a small transparent tube and immersed in a DI water-diluted milk solution (8× dilution of whole milk was added to DI water for optical scattering.) at an 8 mm depth below the US transducer array. The PAUS system imaged this tube at 35 wavelengths (every 5nm from 700nm to 870nm). Then, the identical tube with a 0.4M cupric sulfate solution ($CuSO_4 \cdot 5H_2O$) was immersed into the same DI water-diluted milk solution and positioned at the same location (controlled by B-mode ultrasound) as the tube for tozuleristide, and the PAUS system imaged this second tube at the same 35 wavelengths. (b) The 0.4M $CuSO_4 \cdot 5H_2O$ nonradiative absorption spectrum. The UV-VIS spectrophotometry (BioTek, Epoch 2) was used to measure the nonradiative absorbance spectrum of 0.1M $CuSO_4 \cdot 5H_2O$ from 700nm to 870nm with an increment of 1nm. A concentration of 0.1M was chosen to avoid saturating UV-VIS spectrophotometry measurements. Then, the absorbance was converted to the absoprtion coefficient. Finally, the absorption coefficient of 0.4M $CuSO_4 \cdot 5H_2O$ was obtained as 4× the absorption coefficient of 0.1M $CuSO_4 \cdot 5H_2O$. The absorbance accuracy of the UV-VIS spectrophotometry (BioTek, Epoch 2) is ± 0.01A (A=absorbance). (c) Calculation of PA magnitudes of tozuleristide and $CuSO_4 \cdot 5H_2O$ at 35 wavelengths in the tube experiments. The PA magnitudes of tozuleristide and $CuSO_4 \cdot 5H_2O$ were calculated as the average over the 3×3 pixel region within the yellow box regions. Noise levels were calculated as the average over the 3×3 pixel region within the white box regions and subtracted from the calculated PA magnitudes. (d) The normalized PA magnitudes of tozuleristide and $CuSO_4 \cdot 5H_2O$ at the 35 wavelengths calculated from (c). To calculate the error, the standard deviation (STD) of the pixel values within the white box regions in (c) from all the 35 wavelengths was calculated. Then, the calculated STD was normalized by the same normalization factor as the normalized PA magnitudes of tozuleristide and $CuSO_4 \cdot 5H_2O$, and was equal to 0.0074.

# Supplementary Note 2. Image processing steps for spectroscopic PAUS

## *Fluence compensation*

The wavelength dependence $PA_i(\vec{r},\lambda)$ of the magnitude of the PA signal at position $\vec{r}$ reconstructed from the $i^{th}$ fiber (the total number of fibers surrounding the US probe is $N_{fiber}$) illumination can be represented as

$$PA_i(\vec{r},\lambda) = \Gamma \mu_a^{absorber}(\lambda)\Phi_i(\vec{r},\lambda) , \tag{S3}$$

where $\Gamma$ is the Grüneisen coefficient, $\mu_a^{absorber}(\lambda)$ is the optical absorption coefficient of the absorber at wavelength $\lambda$, and $\Phi_i(\vec{r},\lambda)$ is the optical fluence at position $\vec{r}$ and wavelength $\lambda$ generated by the short laser pulse from the $i^{th}$ fiber.

The total PA image $PA(\vec{r},\lambda)$ at position $\vec{r}$ and wavelength $\lambda$ can be obtained by coherent summation of partial PA images $PA_i(\vec{r},\lambda)$ acquired for each of $N_{fiber} = 20$ fibers:,

$$PA(\vec{r},\lambda) = \sum_{i=1}^{N_{fiber}} PA_i(\vec{r},\lambda) = \Gamma \mu_a^{absorber}(\lambda) \sum_{i=1}^{N_{fiber}} \Phi_i(\vec{r},\lambda) . \tag{S4}$$

To calculate the optical absorption coefficient $\mu_a^{absorber}(\lambda)$ from PA measurements $PA_i(\vec{r},\lambda)$, the optical fluence $\Phi_i(\vec{r},\lambda)$ must be estimated and compensated. In the fast-sweep approach, tissue is illuminated from fibers sequentially (unlike the conventional approach of simultaneous illumination from all the fibers surrounding the probe). Movement of the optical source's position with fiber location can be used to probe the dependence of optical fluence $\Phi_i(\vec{r},\lambda)$ on distance and estimate optical propagation parameters. Thus, unlike other approaches, laser fluence can be estimated directly (as described below) without prior knowledge of tissue optical properties or additional auxiliary measurements.

At depths exceeding a few times the photon transport mean free paths ($l_{free}(\lambda) = 3D(\lambda) = 1/\mu_s'(\lambda)$, where $D(\lambda)$ is the diffusion coefficient of the medium and $\mu_s'(\lambda)$ is the reduced scattering coefficient of the medium), optical fluence $\Phi_i(\vec{r},\lambda)$ formed by illumination from the $i^{th}$ fiber can be described by the diffusion approximation to the light transport equation [2]. As shown in Supplementary Fig. 6 in Jeng et al.[2], the optical source $S_i(\vec{r},\lambda)$ (for the $i^{th}$ fiber) can be approximated as a pencil beam incident on tissue with angle $\theta_i$ with respect to the normal direction of the tissue surface, and the optical fluence can be approximated by the illumination from two isotropic dipole sources [2,3] as

$$\Phi_i(\vec{r},\lambda) = \gamma(\lambda)\left(\frac{1}{4\pi D(\lambda)|\vec{r}-\vec{r_{p,i}}|}e^{-\mu_{eff}(\lambda)|\vec{r}-\vec{r_{p,i}}|} - \frac{1}{4\pi D(\lambda)|\vec{r}-\vec{r_{n,i}}|}e^{-\mu_{eff}(\lambda)|\vec{r}-\vec{r_{n,i}}|}\right) , \tag{S5}$$

where $\mu_a(\lambda)$ is the absorption coefficient of the medium, $\gamma(\lambda)$ is a scaling factor depending on the wavelength, $\mu_{eff}(\lambda) = \sqrt{3\mu_a(\lambda)\mu_s'(\lambda)}$ is the effective light attenuation coefficient of the medium, $\vec{r_{p,i}}$ and $\vec{r_{n,i}}$ are the position vectors of the positive and negative dipole sources for the $i^{th}$ fiber, respectively [2,3], and all other parameters are as defined in Eq. (S3).

When each fiber's optical fluence $\Phi_i(\vec{r},\lambda)$ is normalized by the sum of $\Phi_i(\vec{r},\lambda)$ over all fibers (Eq. (S6)), the only unknown variables in Eq. (S6) are $\mu_{eff}(\lambda)$ and $\mu_s'(\lambda)$ (the measured PA signal is also normalized in Eq. (S7)):

$$\Phi_i^{norm}(\vec{r},\lambda) = \frac{\Phi_i(\vec{r},\lambda)}{\sum_{i=1}^{N_{fiber}} \Phi_i(\vec{r},\lambda)} , \tag{S6}$$

$$PA_i^{norm}(\vec{r},\lambda) = \frac{PA_i(\vec{r},\lambda)}{\sum_{i=1}^{N_{fiber}} PA_i(\vec{r},\lambda)} . \tag{S7}$$

Therefore, the optimal estimate of $\mu_{eff}(\lambda)$ and $\mu_s'(\lambda)$ can be achieved by minimizing the mean squared error between the normalized measured PA signal $PA_i^{norm}(\vec{r},\lambda)$ and optical fluence $\Phi_i^{norm}(\vec{r},\lambda)$ computed from the diffusion equation over all fibers and multiple pixels:

$$\hat{\mu}_{eff}(\lambda), \hat{\mu}_s'(\lambda) = \underset{\mu_{eff}(\lambda) \text{ and } \mu_s'(\lambda)}{\arg\min} \frac{1}{N_{pixel}} \sum_{k=1}^{N_{pixel}} w_k(\lambda)\left(\frac{1}{N_{fiber}} \sum_{i=1}^{N_{fiber}} |PA_i^{norm}(\vec{r_k},\lambda) - \Phi_i^{norm}(\vec{r_k},\lambda)|^2\right) , \tag{S8}$$

where $w_k(\lambda)$ is the weighting factor for the $k^{th}$ pixel at the wavelength $\lambda$, $N_{pixel}$ is the total number of pixels used for optimization, and all other parameters are defined in Eqs. (S6) and (S7). Here we only included pixels with magnitudes exceeding $\langle Noise\_PA \rangle + 3\sigma_{Noise\_PA}$ (where $\langle Noise\_PA \rangle$ is the average noise level, and $\sigma_{Noise\_PA}$ is STD of the noise) in the summation. The weighting factor for any pixel included in the sum is proportional to the square of the PA signal at that pixel averaged over fibers before fluence compensation.

Finally, the fluence-compensated PA image at position $\vec{r}$ and wavelength $\lambda$ after fluence compensation is

$$PA_{fc}(\vec{r},\lambda) = \frac{PA(\vec{r},\lambda)}{\sum_{i=1}^{N_{fiber}} \hat{\Phi}_i(\vec{r},\lambda)} = \frac{\sum_{i=1}^{N_{fiber}} PA_i(\vec{r},\lambda)}{\sum_{i=1}^{N_{fiber}} \hat{\Phi}_i(\vec{r},\lambda)} = k\Gamma\mu_a(\lambda) \ , \tag{S9}$$

where $\hat{\Phi}_i(\vec{r},\lambda)$ is the optimized estimate of the optical fluence at position $\vec{r}$ and wavelength $\lambda$ from Eq. (S8), $k$ is a constant, and all other parameters are as defined in Eq. (S4). Therefore, $PA_{fc}(\vec{r},\lambda)$ is now truly proportional to $\mu_a(\lambda)$ and can be used to represent $\mu_a(\lambda)$.

## *Clutter suppression*

PA signals can be highly corrupted by clutter produced by the laser-ultrasound effect, where strong light absorption at the tissue surface generates laser-ultrasound signals propagating into tissue [4]. These signals reflect from acoustic heterogeneities distributed throughout tissue and are received by the ultrasound transducer as a clutter signal. Note that the clutter signal is generated by ultrasound transmission from the surface (i.e., two-way propagation) but is reconstructed with PA beamforming (i.e., one-way propagation). In our PAUS system, clutter signal patterns differ from those in conventional PAUS systems since the optical source moves across the surface of the transducer from laser firing to laser firing using a discrete fiber array. At every wavelength, the laser beam from each fiber firing can only illuminate the tissue surface region surrounding the fiber so that the surface-generated laser-ultrasound signal will propagate as a narrow beam in the direction perpendicular to the tissue surface and spread as determined by diffraction [5]. The width of the clutter beam does not exceed a few mm in diameter over the frequency range of the receiving US probe (11.3-19.3 MHz with a -6 dB cutoff). Therefore, the clutter signal can only be generated in a narrow tissue region below that fiber. When PA images are reconstructed for each fiber, the beamformed pattern of the clutter signal can be seen to move with fiber firings. Motion of this clutter source provides an opportunity to greatly reduce its influence using different motion-suppression approaches. In this paper we consider only one, compressed averaging [6], which is simple and can be performed in real time.

Compressed averaging is applied to complex PA data (i.e., radio frequency signals after Hilbert transformation to produce IQ data) after fluence compensation. For each wavelength $j$ i.e., [730nm, 744nm, 758nm, 772nm, 786nm, 795nm, 814nm, 828nm, 842nm] in this study), a compression function, such as the root operation of 0.25 (4$^{th}$ order compressed averaging), is applied to the amplitude of IQ data from each fiber $i$ ($i$ an integer changing from 1 to 20) while the phase term remains unchanged:

$$IQ_{compressed}(i,j) = Amplitude\big(IQ(i,j)\big)^{0.25} * Phase(IQ(i,j)) \ , \tag{S10}$$

where $IQ(i,j)$ is the IQ data from the $j^{th}$ wavelength and $i^{th}$ fiber after fluence compensation, and $IQ_{compressed}(i,j)$ is the IQ data from the $j^{th}$ wavelength and $i^{th}$ fiber after applying the compression function.

Then, for each wavelength $j$, compressed IQ data are averaged over all fibers:

$$IQ_{compressed,fiber\ averaged}(j) = \frac{1}{N_{fiber}} \sum_{i=1}^{N_{fiber}} IQ_{compressed}(i,j) \ , \tag{S11}$$

where $IQ_{compressed,fiber\ averaged}(j)$ is the IQ data averaged over all fibers from the $j^{th}$ wavelength, and $N_{fiber}$ is the total number of fibers.

In the final step, for each wavelength $j$, the decompression function (power operation of 4 for the root operation of 0.25 in Eq. (S10)) is applied to the amplitude of $IQ_{compressed,fiber\ averaged}(j)$ while its phase term remains unchanged:

$$IQ_{compressed\ averaging}(j) = Amplitude\big(IQ_{compressed,fiber\ averaged}(j)\big)^4 * Phase(IQ_{compressed,fiber\ averaged}(j)) \ , \tag{S12}$$

where $IQ_{compressed\ averaging}(j)$ is the IQ data after compressed averaging from the $j^{th}$ wavelength.

Finally, the wavelength-compounded PA image after compressed averaging is acquired by coherent sum over all wavelength PA images.

### *Calculation of normalized cross-correlation (NCC) weighting map*

After fluence compensation and compressed averaging, PA signals from all absorbers are present in the wavelength-compounded PA image. To identify only the PA signal from tozuleristide, a normalized cross-correlation (NCC) weighting map is applied [2]. It is obtained by calculating the correlation between the measured PA spectrum after fluence compensation and compressed averaging and the known absorption spectrum of tozuleristide (a high NCC value at a pixel means high confidence that the pixel represents a tozuleristide signal). Specifically, for the $i^{th}$ pixel, the NCC between the measured PA absorption spectrum $S_{PA}(i,\lambda)$ and the reference absorption spectrum of tozuleristide $S_{ref}(\lambda)$ is calculated:

$$NCC(i) = \frac{\sum_{j=1}^{N_\lambda} S_{PA}(i,\lambda_j) S_{ref}(\lambda_j)}{\sqrt{\sum_{j=1}^{N_\lambda} S_{PA}(i,\lambda_j)^2} \sqrt{\sum_{j=1}^{N_\lambda} S_{ref}(\lambda_j)^2}} \quad . \tag{S13}$$

where $\lambda$ is the wavelength, $N_\lambda$ is the total number of wavelengths, $NCC(i)$ is the NCC value for the $i^{th}$ pixel. $S_{PA}(i,\lambda_j)$ and $S_{ref}(\lambda_j)$ denote the PA measured and reference spectrum value at the $j^{th}$ wavelength, respectively.

To identify tozuleristide, a NCC weighting map is calculated pixel-wise by applying a sigmoid function to the calculated pixel-wise NCC values from Eq. (S13). The nonlinear activation in the sigmoid function can suppress pixels with low NCC values and preserve pixels with high NCC values [2]. Therefore, for the $i^{th}$ pixel, the value on the weighting map is

$$Y_{weighting\ map}(i) = \frac{1}{1+e^{-a(NCC(i)-b)}} \quad , \tag{S14}$$

where $Y_{weighting\ map}(i)$ is the value for the $i^{th}$ pixel. $NCC(i)$ is the NCC value for the $i^{th}$ pixel calculated from Eq. (S13), and $a$ denotes the slope of the sigmoid function and $b$ denotes the NCC value at the level of $Y_{weighting\ map} = 0.5$. Because a positive-definite signal is used to compute the NCC, it is highly compressed to a range of values between about 0.9 and 1.0. Given this compression, we selected the values of $a = 300$ and $b = 0.978$ for the NCC-weighting results in Study 2 in the main text to reasonably minimize noise for low NCC regions and highlight high NCC regions.

Finally, the NCC-weighted PA image is generated by multiplying the wavelength-compounded PA image after compressed averaging with the NCC weighting map. Note that the images presented in Fig. 9c and Fig. 9e are not highly sensitive to the specific choice of a and b over a wide range as long as the sigmoid was confined to the compressed range of NCC values.

## References

1    Pelivanov, I. *et al.* Molecular fingerprinting of nanoparticles in complex media with non-contact photoacoustics: Beyond the light scattering limit. *Scientific Reports* **8**, 14425 (2018).
2    Jeng, G.-S. *et al.* Real-time interleaved spectroscopic photoacoustic and ultrasound (PAUS) scanning with simultaneous fluence compensation and motion correction. *Nature communications* **12**, 716 (2021).
3    Kim, M., Jeng, G.-S., O'Donnell, M. & Pelivanov, I. Correction of wavelength-dependent laser fluence in swept-beam spectroscopic photoacoustic imaging with a hand-held probe. *Photoacoustics* **19**, 100192 (2020).
4    Jaeger, M., Bamber, J. C. & Frenz, M. Clutter elimination for deep clinical optoacoustic imaging using localised vibration tagging (LOVIT). *Photoacoustics* **1**, 19-29 (2013).
5    M. B. Vinogradova, O. V. Rudenko and A. P. Sukhorukov, "Wave Theory," (Moscow: Nauka)    p. 383 (1979).
6    Polichetti, M., Varray, F., Béra, J.-C., Cachard, C. & Nicolas, B. A nonlinear beamformer based on p-th root compression—Application to plane wave ultrasound imaging. *Applied Sciences* **8**, 599 (2018).